\documentclass[10pt,twocolumn,letterpaper]{article}

\usepackage{cvpr}              

\usepackage{graphicx}
\usepackage{multirow}
\usepackage{adjustbox}
\usepackage{float}
\usepackage{algorithm}
\usepackage{algpseudocode}
\usepackage{algcompatible}
\usepackage{mathtools}
\usepackage{amsmath} 
\usepackage{amsthm}
\usepackage{siunitx}
\usepackage{colortbl}
\usepackage{xcolor}
\usepackage{balance}
\usepackage{amstext} 
\usepackage{array}   
\usepackage{xspace}
\usepackage{pifont}

\usepackage[accsupp]{axessibility}  

\definecolor{myPink}{RGB}{220,80,140}
\usepackage[
    pagebackref,
    breaklinks,
    colorlinks,
    linkcolor=cvprblue,
    citecolor=cvprblue,
    filecolor=cvprblue,
    urlcolor=myPink
]{hyperref}

\definecolor{cvprblue}{rgb}{0.21,0.49,0.74}
\definecolor{myPink}{RGB}{220,80,140}

\def\paperID{}
\def\confName{CVPR}
\def\confYear{2026}

\newcommand{\ours}[1]{CATRF}

\title{CATRF: Codec-Adaptive TriPlane Radiance Fields for Volumetric Content Delivery}

\author{
Tung-I Chen \qquad
Lingdong Wang \qquad
Subhransu Maji \qquad
Ramesh K. Sitaraman\\
University of Massachusetts Amherst\\
\small \href{https://tung-i.github.io/catrf-cvpr-findings-2026/}{\textcolor{myPink}{\texttt{https://tung-i.github.io/catrf-cvpr-findings-2026/}}}
}

\begin{document}
\maketitle

\begin{abstract}
Volumetric media promises next-generation content delivery applications, but its bandwidth demand remains a key bottleneck. Implicit and hybrid volumetric representations reduce model sizes, yet still require careful coding to reach 2D video-like bitrates. We present \ours{}, a \emph{standard-codec-in-the-loop} compression framework for plane-factorized radiance fields. 
During training, we quantize and pack 2D feature planes into codec-friendly canvases, run a standard codec round trip (JPEG/VP9/HEVC/AV1), then unpack and dequantize the decoded features before volume rendering. We use a straight-through estimator (STE) to insert the non-differentiable, standard codec pipeline into the training loop, allowing radiance-field features to adapt directly to the real, client-side codec distortions without introducing any learned codec parameters.
On both static and dynamic benchmarks, \ours{} consistently achieves a better rate-distortion trade-off over codec-agnostic and learned-codec-in-the-loop baselines, and also outperforms recent compressed 3DGS methods in both compression efficiency and decoding speed.
These results highlight a practical path toward low-bitrate, compression-resilient volumetric representations for free-viewpoint video streaming.
\end{abstract}

\section{Introduction}
\label{sec:intro}

\begin{figure}[t]
  \centering

   \includegraphics[width=1.0\linewidth]{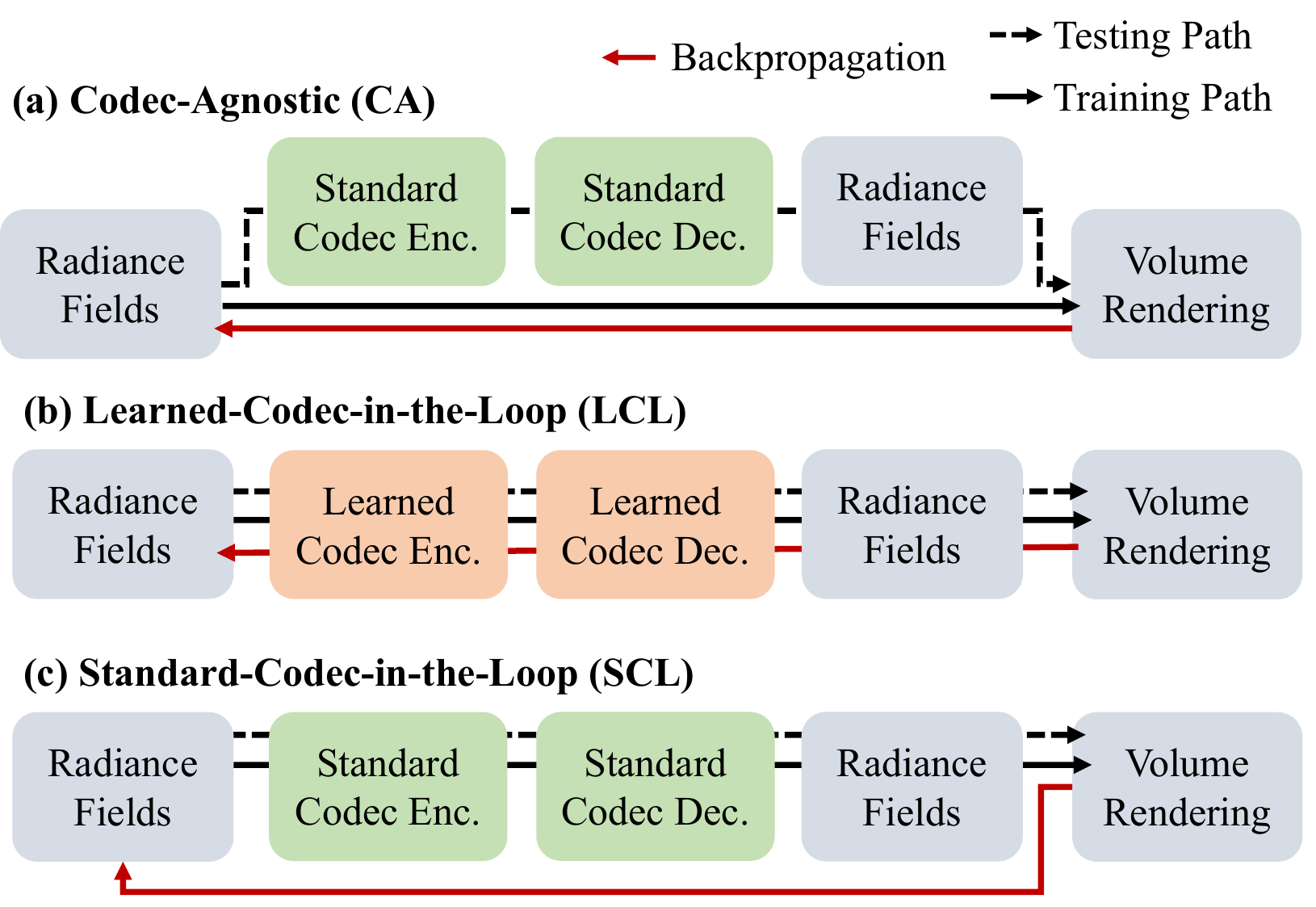}

   \caption{
    Overview of codec-integrated NeRF compression pipelines.
    (a) Codec-agnostic (CA) approaches optimize plane-factorized NeRFs without exposure to codec-induced distortions; when compressed afterward, codec artifacts disrupt feature semantics and degrade quality at low bitrates. (b) Learned-codec-in-the-loop (LCL) methods insert a differentiable learned codec into training to optimize the rate-distortion (RD) trade-off end-to-end. However, they require shipping a content-specific decoder and running ML inference on the client to reconstruct NeRF models. (c) Our standard-codec-in-the-loop (SCL) approach combines the practicality of CA with the benefits of LCL by executing standard codec round trips during training and passing gradients through a straight-through estimator (STE).
    }
   \label{fig:scheme}
\end{figure}

\begin{figure*}[t]
  \centering
   \includegraphics[width=1.0\linewidth]{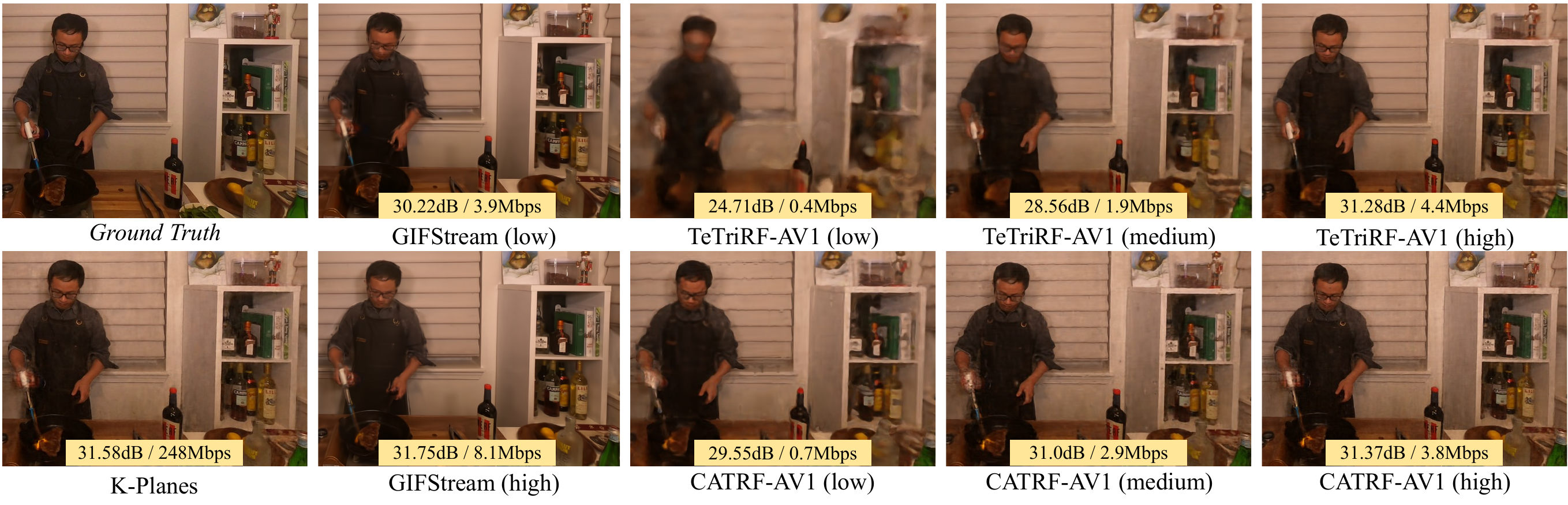}
    
    \caption{
    Qualitative comparisons on the Neural 3D Video benchmark. At high bitrate, both CA and SCL methods produce comparable visual quality. However, as compression becomes more aggressive, the CA baseline exhibits severe blurring and artifacts, whereas \ours{} preserves sharper details and cleaner geometry. Compared with the 3DGS baseline, \ours{} also achieves comparable or better visual quality at substantially lower bitrate. The bitrate is computed at 30 FPS.
    }
   \label{fig:catrf_video_qualitative}
\end{figure*}

Modern 2D video compression and streaming standards are mature and widely deployed, serving high-quality videos at scale and forming the backbone of today’s content platforms~\cite{sodagar2011mpeg, bentaleb2022low}. However, the dynamic world we perceive is inherently 3D rather than 2D. Beyond conventional videos, there is growing interest in free-viewpoint media, where users can view a scene at any time and from any viewpoint. Volumetric content delivery~\cite{bentaleb2025solutions, collet2015high, jin2023capture}, if realized, has the potential to transform content platforms by enabling an immersive viewing experience, such as 3D video playback, free-viewpoint sports, and telepresence.
Delivering 3D content is bandwidth-hungry. For example, Microsoft’s Holoportation reports Gbps-scale rates for streaming a full 3D human~\cite{orts2016holoportation}. Explicit representations such as point clouds and 3D Gaussian Splatting (3DGS) render efficiently but store many per-point attributes. Even with state-of-the-art compression techniques, streaming these representations often remains in the hundreds of Mbps~\cite{chen2024hac, niedermayr2024compressed, wu20244d, shaw2024swings}, whereas today’s streaming systems typically budget only tens of Mbps per user. In contrast, implicit or hybrid representations (e.g., NeRFs and TriPlane~\cite{mildenhall2021nerf, chan2022efficient, chen2022tensorf}) encode scenes as compact neural features. When factorized and coded properly, storage per frame can drop to tens of KB~\cite{wu2024tetrirf, hu2025vrvvc}.
Recent work shows that the rate-distortion (RD) trade-off of NeRF-based representations can be improved by integrating a learned, differentiable codec into the training objective~\cite{li2024nerfcodec, chen2024far, lee2024ecrf, zhang2024rate, hu2025vrvvc}. These learned-codec-in-the-loop (LCL) frameworks employ a learned entropy model or neural codec to encode neural features into bitstreams, which are then used at decoding time to reconstruct the 3D representation. Such frameworks are effective at reducing feature bitrate while preserving reconstruction quality. However, when codec parameters are trained per scene, they must also be transmitted to the client for decoding, akin to neural-enhanced video streaming where ML models are shipped with the content~\cite{yeo2018neural, yeo2020nemo}. A content-specific decoder must then be distributed and maintained on devices, incurring non-negligible bandwidth overhead and deployment complexity (see ~\cref{tab:memory_breakdown}).
Another line of work converts 3D representations into 2D signals that standard codecs compress extremely well. For instance, MPEG’s video-based point cloud compression (V-PCC) projects point attributes to patch videos, enabling HEVC/AV1 coding and delivery through adaptive bitrate (ABR) streaming standards~\cite{graziosi2020overview, sodagar2011mpeg, sani2017adaptive}. Inspired by this 3D-via-2D idea, recent NeRF compression methods use factorized tensor planes as the representation to operate on 2D feature maps~\cite{wu2024tetrirf, wang2024videorf, wang2024v, lee2025compression}. In these codec-agnostic (CA) compression pipelines, feature planes are optimized with spatial-temporal consistency losses and compressed afterward with off-the-shelf codecs. While being hardware- and ABR-friendly, CA representations degrade significantly at low bitrates because feature planes never see the real codec-induced distortions during optimization and therefore remain vulnerable to quantization, block transforms, and in-loop filtering (see \cref{fig:n3dv_rd_curve} and \cref{fig:catrf_video_qualitative}).
We present Codec-Adaptive TriPlane Radiance Fields (\ours{}), the first standard-codec-in-the-loop (SCL) training framework for plane-factorized NeRFs. A prevailing assumption in codec-integrated training is that the codec itself must be differentiable. Accordingly, prior work either replaces a standard codec with a differentiable proxy that approximates codec behavior~\cite{isik2023sandwiched, li2024neural, jia2025towards} or jointly optimizes a neural encoder-decoder pair so gradients can flow end-to-end~\cite{li2024nerfcodec, kang2025codecnerf}. \ours{} overturns this assumption: instead of making the codec differentiable, we treat the real VP9/HEVC/AV1 round trip as a black box and use a straight-through estimator (STE) to pass gradients through the full encode-decode process. Specifically, we quantize and pack multi-channel appearance planes and the density grid into 2D canvases before the codec round trip, then render from the decoded features when computing reconstruction loss. To optimize RD trade-off, we regularize features with spatial total-variation and temporal-consistency terms that encourage smoother, codec-friendly patterns. We further introduce a caching mechanism that reuses decoded planes across steps, avoiding per-minibatch codec round trips and substantially improving training efficiency.
We evaluate \ours{} on both static and dynamic 3D benchmarks with diverse baselines. On static scenes, \ours{}-JPEG achieves a $43\%$ BD-rate reduction over state-of-the-art LCL compression methods. On dynamic scenes, \ours{} paired with VP9/HEVC/AV1 consistently outperforms codec-agnostic TeTriRF~\cite{wu2024tetrirf} across different codec configurations, with the largest gains at low bitrates. Beyond NeRF baselines, \ours{} also outperforms the state-of-the-art compressed 3DGS methods: it achieves $17\%$ bitrate savings over HAC++~\cite{chen2025hac++} and $39\%$ BD-rate reduction over GIFStream~\cite{li2025gifstream}, while providing substantially faster reconstruction of 3D representations from the encoded bitstreams through standard-codec decoding (see \cref{tab:decoding_speed}). 

\begin{figure}[t]
  \centering

   \includegraphics[width=1.0\linewidth]{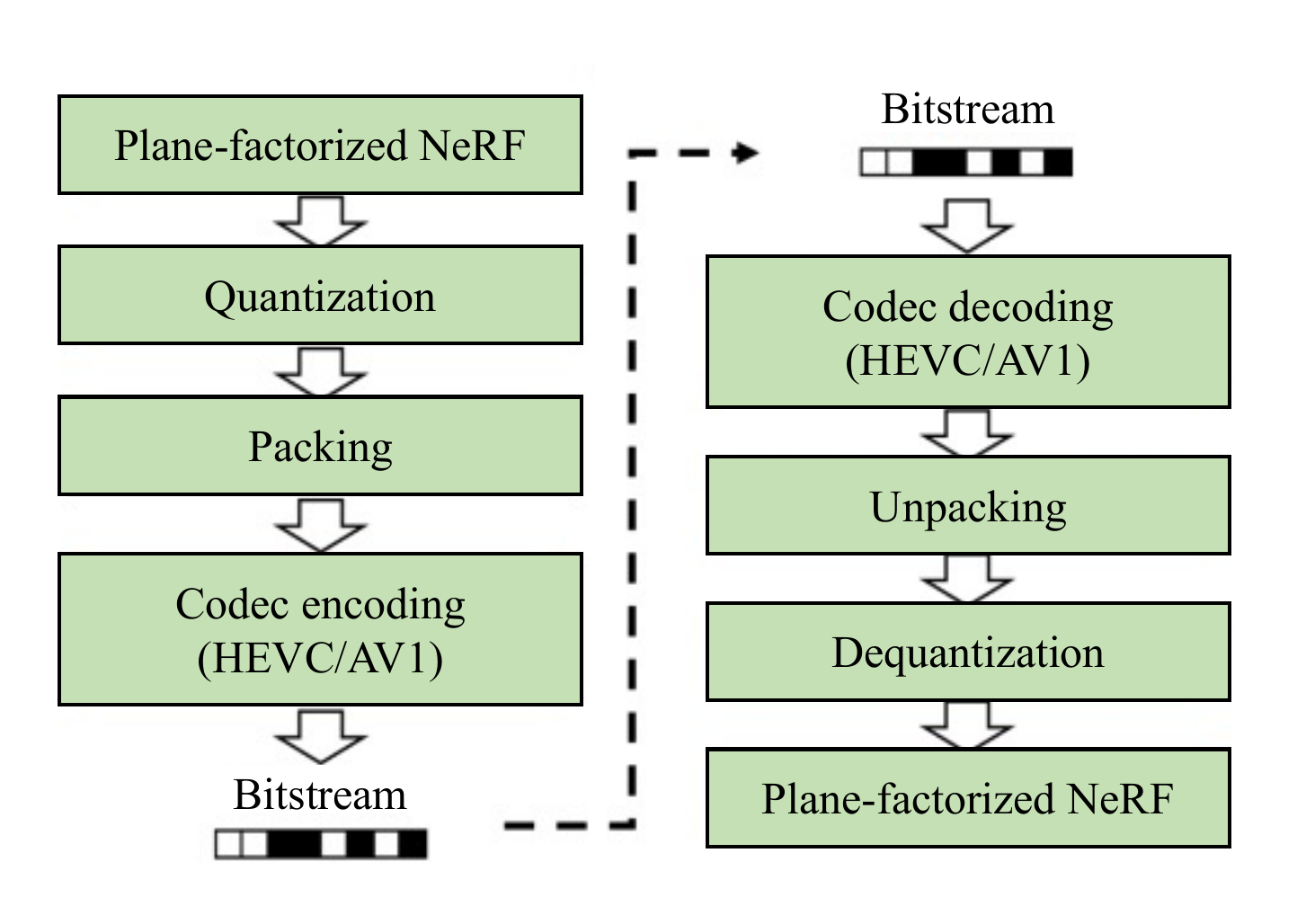}

   \caption{
   Illustration of the encode–decode codec round trip. In a streaming setting, the client receives and renders the \emph{decoded} NeRF model reconstructed from the encoded bitstream.
   }
   \label{fig:eval_workflow}
\end{figure}

To summarize, our key contributions are:
\begin{itemize}[leftmargin=1.2em,topsep=2pt,itemsep=2pt]
    \item We present the first standard-codec-in-the-loop training framework for plane-factorized radiance fields, where neural features are optimized in the decoded space produced by a real codec. Our design is practical: it works with off-the-shelf JPEG/VP9/HEVC/AV1 encoders/decoders, and requires no learned neural codecs or learned entropy-models.
    \item We study key design choices for neural feature compression, including quantization, packing layouts, and a caching strategy that amortizes codec round-trip overhead during training.
    \item Through extensive experiments on both static and dynamic benchmarks, we show that \ours{} achieves up to $63\%$ BD-rate reduction over state-of-the-art NeRF compression methods and outperforms state-of-the-art 3DGS baselines in both bitrate savings and decoding speed.
\end{itemize}

\section{Related Work}
\label{sec:related}

\subsection{Memory-Efficient Volumetric Representations}
A major direction in compact 3D scene representation is to encode scenes into lightweight neural features together with a small decoder MLP. TriPlane Radiance Fields employ a hybrid explicit–implicit representation that represents a scene with three orthogonal feature planes~\cite{chan2022efficient, muller2022instant, sun2022direct, fridovich2022plenoxels}. TensoRF compresses neural volumes via matrix–vector factorizations~\cite{chen2022tensorf}; K-Planes extends plane factorization across space–time to exploit temporal redundancy~\cite{fridovich2023k}; BiRF further improves rate-distortion by binarizing INGP hash embeddings~\cite{muller2022instant, shin2023binary}.
A parallel line of work studies compressed 3D Gaussian Splatting (3DGS). Static 3DGS can be compressed by encoding anchor-based Gaussian attributes with compact latent features~\cite{chen2024hac, chen2025hac++}. For dynamic 3DGS, the number of parameters can be further reduced by exploiting the residual information between frames~\cite{li2024spacetime,li2025gifstream, girish2024queen, sun20243dgstream}. 
However, reducing parameter count alone does not guarantee a low-bitrate streaming representation. Practical payloads can still remain on the order of megabytes per frame. This motivates subsequent work that couples compact volumetric representations with explicit coding modules for more bitrate savings~\cite{hu2025vrvvc, zhang2024rate, zheng2024jointrf}.

\subsection{Neural Feature Compression}
Compared with other volumetric representations, TriPlane features have an inherent 2D structure, making them compatible with standard 2D neural networks and mature image-processing pipelines~\cite{shue20233d, li2024instant3d, gupta20233dgen, li2024nerfcodec, lee2025compression}. 
TeTriRF~\cite{wu2024tetrirf} shows that feature planes can be compressed with standard video codecs to achieve compact storage ($10$-$100$ KB/frame), suggesting a promising \textit{features-as-video} paradigm. However, neural features are much more sensitive than natural images to codec distortions. When training does not account for reconstruction errors introduced by the codec-induced distortions, low-bitrate compression causes severe rendering degradation.
A complementary line of work replaces standard codecs with learned differentiable codecs to enable codec-in-the-loop optimization~\cite{li2024nerfcodec, hu2025vrvvc, chen2024far, zheng2024jointrf}. CNC~\cite{chen2024far} learns a context model over 3D hash embeddings for entropy reduction, while NeRFCodec~\cite{li2024nerfcodec} learns content-specific adaptors around a pre-trained neural codec. 
However, when an entropy model or neural codec is trained per scene, its parameters must also be transmitted for decoding and should therefore be counted in the total bandwidth budget. Moreover, hardware support for neural decoding remains much less ubiquitous than that of standard codecs on consumer devices, resulting in limited practicality in real-world applications.

\subsection{Straight-Through Estimators}
The Straight-Through Estimator (STE) enables training with discrete or non-differentiable operations by using the true operation in the forward pass and an approximate gradient in the backward pass~\cite{bengio2013estimating}. In the field of neural-based compression, it is widely used to bypass quantization and any non-differentiable components during optimization~\cite{courbariaux2015binaryconnect, cai2019proxylessnas, balle2017iclr, balle2018tisl}. 
For codec-aware training, prior work usually relies on a differentiable \textit{virtual codec}, where prediction, transform, and entropy coding are modeled by custom surrogate modules instead of running a real standard codec in the loop. This design appears broadly in neural codec literature, such as sandwiched video compression~\cite{isik2023sandwiched} and the DCVC family of neural codecs ~\cite{li2024neural,jia2025towards}. 
By contrast, we execute a full off-the-shelf codec (HEVC/AV1) inside training and use STE to backpropagate through the entire real encode-decode process. This makes \ours{} the first standard-codec-in-the-loop framework for plane-factorized radiance fields to optimize through a real deployed codec rather than a differentiable proxy.

\begin{figure*}[t]
  \centering

   \includegraphics[width=0.9\linewidth]{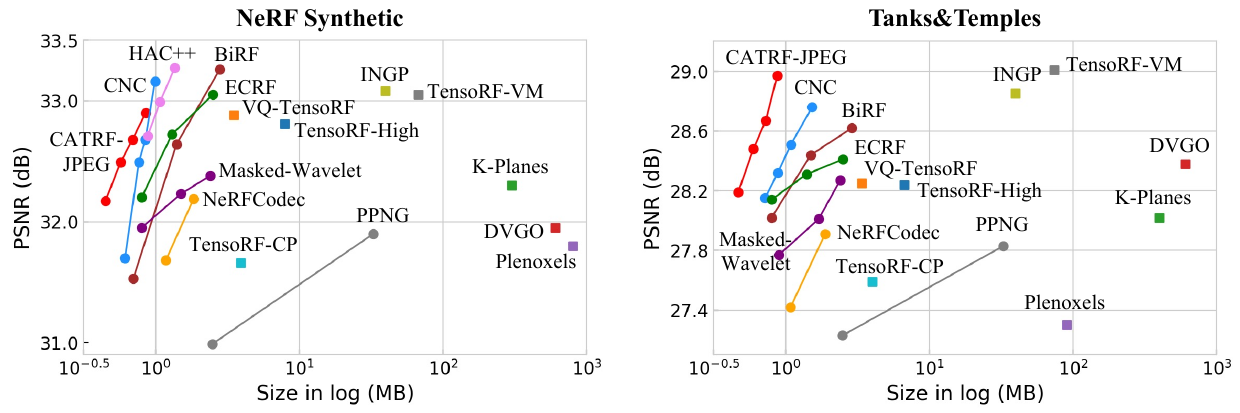}

    \caption{
    Comparison with baselines on the NeRF Synthetic (left) and Tanks and Temples (right) benchmarks. On NeRF Synthetic, \ours{}-JPEG achieves average bitrate savings of $23\%$ and $17\%$ at matched PSNR over CNC~\cite{chen2024far} and HAC++~\cite{chen2025hac++}, respectively.
    }
   \label{fig:catrf_image_quantitative}
\end{figure*}

\begin{figure*}[t]
  \centering

   \includegraphics[width=0.96\linewidth]{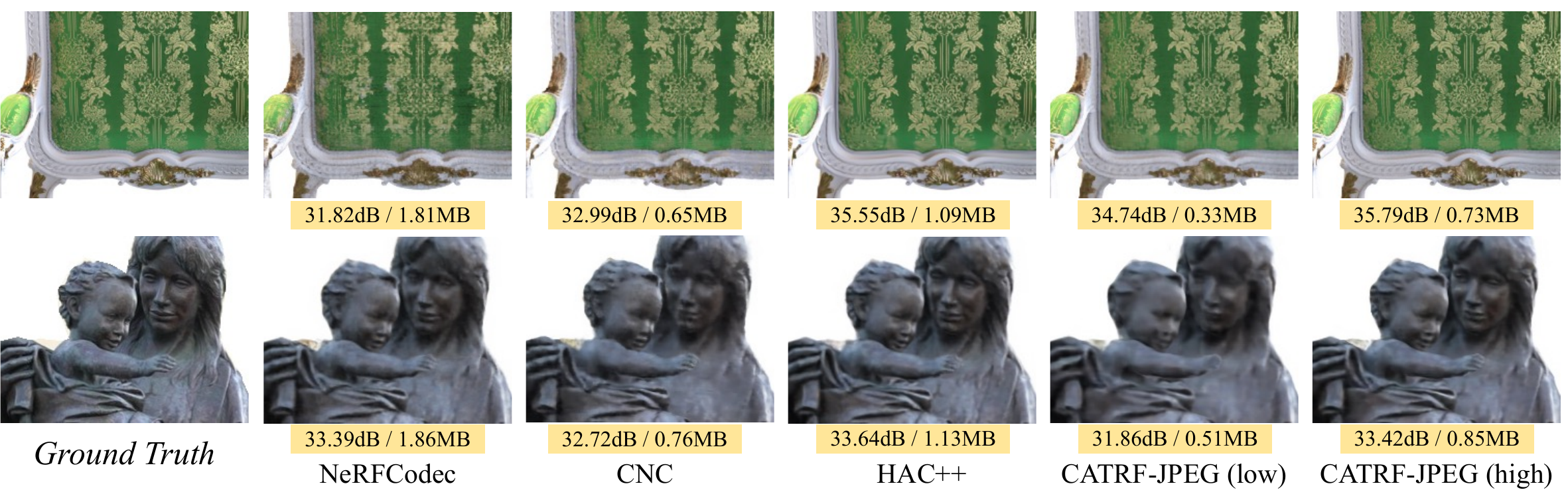}

    \caption{
    Qualitative comparisons on \textit{Chair} from NeRF Synthetic and \textit{Family} from Tanks and Temples. Compared with recent learned-codec and 3DGS compression baselines, \ours{}-JPEG offers a flexible operating range: it can achieve substantially lower bitrate with only modest quality loss, or deliver sharper details and fewer artifacts at a slightly higher rate.
    }
   \label{fig:catrf_image_qualitative}
\end{figure*}
\section{Method}
\label{sec:method}

Our goal is simple: \textit{make the decoded radiance field features that the client will actually render be the ones optimized during training}. 
In each training step, we (i) quantize and pack plane-factorized radiance-fields into 2D canvases (i.e, images with integer pixel values), (ii) run a codec round trip, encoding and decoding the canvases, (iii) unpack and dequantize the reconstructed features, and (iv) render the radiance-fields to compute the reconstruction loss. 
Unlike previous LCL frameworks which are restricted to differentiable encoders and decoders, we propose to pass gradients across the entire codec roundtrip using a straight-through estimator (STE), ensuring no neural decoders or entropy model parameters must be shipped for client-side decoding.

\subsection{Background and Notation}
\label{subsec:bg_notation}

Many recent NeRF variants represent a scene with plane-factorized features and an explicit 3D density grid for efficient rendering~\cite{chan2022efficient, chen2022tensorf, sun2022direct, fridovich2023k}. This design is attractive for compression, since the Euclidean structure of these tensors maps naturally to image/video codecs.
We denote the three orthogonal feature planes of TriPlane radiance fields by $\mathcal{P}=\{P^{xy},P^{xz},P^{yz}\}$ with $P^{ax}\in\mathbb{R}^{C\times H\times W}$, where $ax\in [xy, xz, yz]$ and $C$ is the channel dimension.
The 3D density grid is denoted by $D\in\mathbb{R}^{D_y\times D_x\times D_z}$.
We denote the lightweight neural renderer that decodes aggregated features and view direction into color and density by $f_\phi$.
For a camera pose $\pi$, we write the rendering call as $\mathcal{R}(\mathcal{P},D,\pi;\phi)$, which produces a rendered image $I$.
In a streaming setting, the data that must be transmitted include the encoded radiance-field bitstreams, the neural renderer weights $\phi$, and any parameters required to retrieve $(\mathcal{P},D)$.
%


\subsection{Volume Rendering}
\label{subsec:plane_factored}
Plane factorization stores features on axis-aligned 2D planes and queries them by bilinear sampling $\mathrm{bi}(\cdot)$. For a 3D location $\mathbf{x} = (x,y,z)$, we define its 2D projections as $\mathbf{u} = (x,y)$, $\mathbf{v} = (x,z)$, and $\mathbf{w} = (y,z)$. The tri-plane feature at $\mathbf{x}$ is
\begin{align*}
\mathbf{f}_a(\mathbf{x}) = \big[
  &\mathrm{bi}(P^{xy}, \mathbf{u}),\;
   \mathrm{bi}(P^{xz}, \mathbf{v}),\;
   \mathrm{bi}(P^{yz}, \mathbf{w})
\big],
\end{align*}
where the three sampled vectors are aggregated (e.g., by summation or concatenation) into a single feature vector $\mathbf{f}_a(\mathbf{x})$.
The density feature is obtained by trilinear sampling from the 3D grid, $s(\mathbf{x}) = \mathrm{trilinear}(D, \mathbf{x})$.

Given a view direction $\mathbf{v}_{\text{view}}\in\mathbb{R}^3$, the renderer $f_\phi$ decodes per-sample density and color:
\[
\sigma(\mathbf{x}),\,\mathbf{c}(\mathbf{x},\mathbf{v}_{\text{view}})
\;=\;
f_\phi\!\big(\mathbf{f}_a(\mathbf{x}),\, s(\mathbf{x}),\, \mathbf{v}_{\text{view}}\big),
\]
where $\sigma(\mathbf{x})$ is the volume density and $\mathbf{c}(\mathbf{x},\mathbf{v}_{\text{view}})\in\mathbb{R}^3$ is the view-dependent RGB color at location $\mathbf{x}$. For a camera ray $\mathbf{r}(t)=\mathbf{o}+t\,\mathbf{d}$ with samples $\{t_k\}_{k=1}^{K}$ and $\Delta_k=t_{k+1}-t_k$, we evaluate $(\sigma_k,\mathbf{c}_k)$ at 3D locations $\mathbf{x}_k=\mathbf{r}(t_k)$ using $(\mathcal{P},D)$ and $f_\phi$. Given that $\alpha_k = 1 - \exp(-\sigma_k \Delta_k)$, the final pixel color $\mathbf{C}$ can be computed by alpha compositing:
\begin{align*}
T_k = \exp\!\Big(-\sum_{j<k} \sigma_j \Delta_j\Big), 
\mathbf{C} = \sum_{k=1}^{K} T_k\,\alpha_k\,\mathbf{c}_k.
\end{align*}

\begin{figure*}[t]
  \centering
   \includegraphics[width=1.0\linewidth]{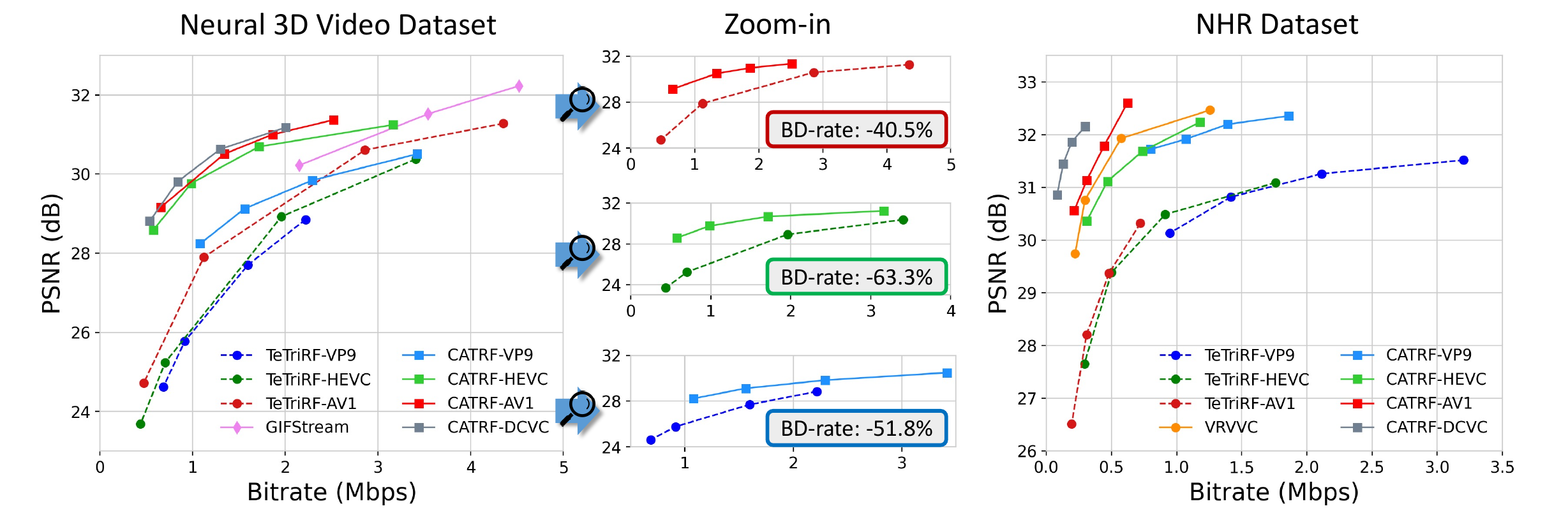}
   \caption{
    Rate-distortion (RD) curves. We compare \ours{} with codec-agnostic TeTriRF~\cite{wu2024tetrirf} across different codecs and target bitrates. \ours{} consistently outperforms TeTriRF on both benchmarks with substantial BD-rate reductions. \ours{}-AV1 also outperforms the state-of-the-art dynamic 3DGS baseline, GIFStream~\cite{li2025gifstream}, achieving $39\%$ BD-rate savings.
    }
   \label{fig:n3dv_rd_curve}
\end{figure*}

\subsection{Quantization and Packing}
\label{subsec:quant-pack}

Standard codecs accept fixed-precision image/video inputs (e.g., 8/10/12-bit integers) in specific pixel formats, whereas the factorized feature planes of NeRF models are multi-channel, floating-point tensors. 
We therefore have to quantize features into a normalized range $[0,1]$ and pack multi-channel tensors into 2D canvases with either 1 or 3 channels (grayscale or RGB). In implementation, canvases in $[0,1]$ are linearly mapped to 8-bit integers before encoding and converted back to $[0,1]$ after decoding. Let $X$ denote a tensor plane from the radiance fields, and consider two schemes, \textit{AbsMax} and \textit{Channelwise} to map floating-point tensors to $[0,1]$. For \textit{AbsMax} (absolute max quantization),
we use a pre-defined, fixed range $[\alpha,\beta]$ shared by all channels. For \textit{Channelwise}, we estimate a per-channel central 95\% range for each channel, and use the resulting lower and upper bound $[\alpha,\beta]$ of each channel to perform channel-wise quantization.
\begin{align*}
X_{01} &= \mathrm{clip}\!\left(\frac{X-\alpha}{\beta-\alpha},\,0,1\right)
&& \text{(quantization)}\\
\widehat{X} &= X_{01}(\beta-\alpha)+\alpha
&& \text{(dequantization)}.
\end{align*}


Standard image/video codecs, however, operate on 1- or 3-channel images, so we have to pack the multi-channel tensor $X_{01}$ into a codec-friendly canvas $Y = \mathsf{pack}(X_{01})$ with $Y\in \mathbb{R}^{C_{\text{out}} \times H' \times W'}$ and $C_{\text{out}} \in \{1,3\}$. We consider three packing strategies for appearance planes $\mathcal{P}$. Take $[C=48, H, W]$ as an example. \noindent\textit{FlattenGray} (no cross-channel mixing) stacks all channels spatially into a single monochrome canvas, i.e., $[1,\,6H,\,8W]$; \noindent\textit{FlattenRGB} (grouped channels) arranges channels into three groups of 16, resulting in a layout $[3,\,4H,\,4W]$; \noindent\textit{PixelShuffle} (space-to-depth) rearranges channels into spatial neighborhoods, resulting in $[3,\,4H,\,4W]$ as well.
We provide an ablation study in \cref{tab:ablation} to compare the impact of each packing choice. For the density grid $D$ that directly controls opacity, we tile it into a single monochrome canvas and encode it as grayscale to avoid mixing across channels.





\subsection{Caching for Efficient Training}
Given the three appearance canvases and the density monochrome canvas, we invoke a standard codec on each canvas to encode it into a separate bitstream and then decode it back to simulate client-side decoding in a streaming scenario:
\[
(\widehat{Y},\, b) = \mathcal{C}_q(Y),
\]
where $\mathcal{C}_q(.)$ denotes the codec round trip operation (i.e., encode-then-decode); $b$ is the encoded bit count and $q$ is the codec parameters for rate-control, such as a quantization parameter (QP). After unpacking and dequantization, the reconstructed features $(\widehat{\mathcal{P}}, \widehat{D})$ are fed to the renderer. Since naively invoking $\mathcal{C}_q$ on every batch would dominate training time, we propose to maintain a cache of the decoded $(\widehat{\mathcal{P}}, \widehat{D})$. The cache is refreshed only when (i) a fixed number $M$ of global steps has elapsed, or (ii) the current frame has drifted too far from the snapshot stored at the last refresh, determined by a small threshold $\epsilon$. When the cache is up to date, we reuse the cached $(\widehat{\mathcal{P}}, \widehat{D})$ to render and compute the reconstruction loss. The detailed caching algorithm is presented in \cref{sec:ste_cache} of the supplementary.


\subsection{Standard-Codec-in-the-Loop (SCL)}
\label{subsec:codec-loop}

Prior codec-in-the-loop frameworks rely on differentiable codec proxies or learned codecs/entropy models to pass gradients~\cite{li2024nerfcodec, chen2024far}. Our approach removes this restriction by employing a straight-through estimator (STE) to bypass the entire codec roundtrip during backpropagation, which allows us to insert standard, non-differentiable codecs (JPEG/HEVC/AV1) directly into the training loop of neural radiance fields.
We define a stop-gradient operator $\texttt{detach}(\cdot)$ that is identical to the identity in the forward pass but blocks gradients, i.e., $\texttt{detach}(X)=X$ and $\frac{\partial}{\partial X}\texttt{detach}(X)=\mathbf{0}$. Let $\widehat{P}^{ax}$ and $\widehat{D}$ denote the reconstructed feature planes and density grids. We form the STE overrides as:
\begin{align*}
\widetilde{P}^{ax} &= \widehat{P}^{ax} + \big(P^{ax} - \texttt{detach}(P^{ax})\big), \\
\widetilde{D}      &= \widehat{D}      + \big(D      - \texttt{detach}(D)\big).
\end{align*}
The forward pass uses $(\widetilde{\mathcal{P}}, \widetilde{D})$ to render images, which are numerically equal to the decoded features $(\widehat{\mathcal{P}}, \widehat{D})$, but in backpropagation we have
\[
\frac{\partial \widetilde{P}^{ax}}{\partial P^{ax}} = \mathbf{I},\qquad
\frac{\partial \widetilde{D}}{\partial D} = \mathbf{I},
\]
so gradients bypass the codec and flow directly to the pre-codec features $(\mathcal{P}, D)$. We evaluate several gradient-surrogate choices for propagating gradients through the non-differentiable codec round trip, including vanilla STE~\cite{bengio2013estimating}, modified STE~\cite{mack2025efficient}, and a hybrid STE+SPSA estimator~\cite{spall1998overview}. The gradient-surrogate ablation is presented in \cref{tab:grad_surrogate_ablation} of the supplementary, which shows STE provides the best robustness. The pseudo-code below summarizes each training step of the proposed SCL pipeline:
\begin{enumerate}
\item For each axis $ax\in\{xy,xz,yz\}$:
  \begin{enumerate}
  \item $X_{01}\leftarrow \mathsf{Q}(P^{ax})$;\quad $Y\leftarrow \mathsf{pack}(X_{01})$.
  \item $(\widehat{Y},\mathrm{bits})\leftarrow \mathcal{C}_q(Y)$;\quad $\widehat{P}^{ax}\leftarrow \mathsf{Q}^{-1}(\mathsf{unpack}(\widehat{Y}))$.
  \item $\widetilde{P}^{ax}\leftarrow \widehat{P}^{ax} + \big(P^{ax}-\texttt{detach}(P^{ax})\big)$.
  \end{enumerate}
\item Process density via quantization and mono tiling. Then, \\ $\widetilde{D}\leftarrow \widehat{D} + \big(D-\texttt{detach}(D_t)\big)$.
\item Render: $I\leftarrow \mathcal{R}(\widetilde{\mathcal{P}},\widetilde{D},\pi;\phi)$; compute $\mathcal{L}$ and update $(\mathcal{P},D,\phi)$.
\end{enumerate}
%


\subsection{Loss Functions}
To optimize \ours{} for a given camera $\pi$ and the ground-truth image $I^*$, we define the total objective as:
\[
\mathcal{L} = \lambda_{\text{rec}}\,
\underbrace{\big\| \mathcal{R}(\widetilde{\mathcal{P}},\widetilde{D},\pi;\phi)-I^* \big\|_1}_{\text{image reconstruction}}
\;+\;
\lambda_{\text{tv}}\,
\underbrace{\big(\mathrm{TV}(\mathcal{P})+\mathrm{TV}(D)\big)}_{\text{spatial smoothness for bitrate}}.
\]

This total-variation~\cite{chen2022tensorf} encourages spatial smoothness of the packed canvases, reducing high-frequency residuals that would otherwise inflate the bitrate. For dynamic scenes, we additionally incorporate intra-group and inter-group regularization as proposed in~\cite{wu2024tetrirf} to ensure temporal consistency. Unlike previous compression methods that require a rate-loss to optimize the RD trade-off of their learned entropy models~\cite{zheng2024jointrf, zhang2024rate, hu2025vrvvc}, \ours{} does not employ an explicit rate term. This design choice stems from the fact that our entropy model and rate-control strategies are well-established by \textit{fixed, industry-standard codecs}. These codecs have been optimized over decades for near-optimal RD trade-off. Consequently, minimizing reconstruction distortion at a fixed standard codec configuration implicitly drives the features toward the most efficient point of the RD curve. In our experiments, adding additional rate-related losses does not provide measurable gains.





\section{Experiments}
\label{sec:experiments}

\begin{table}[t!]
    \setlength{\tabcolsep}{4pt}  
    \centering
    \resizebox{\linewidth}{!}{
    \begin{tabular}{ccccccccccc}
    \toprule
    & PSNR$_\uparrow$ & SSIM$_\uparrow$ & LPIPS$_\downarrow$ & Size$_\downarrow$ & Bitrate$_\downarrow$  \\
    & & & (VGG) & (MB) & (Mbps) \\
    \midrule
    K-Planes~\cite{fridovich2023k}  & 31.58  & 0.941 & 0.14 & 1.0  & 248  \\
    StreamRF~\cite{li2022streaming}  & 30.68  &- & - & 31.4  & 7536 \\
    NeRFPlayer~\cite{song2023nerfplayer} & 30.29  & 0.910 & 0.31 & 17.1  & 4104 \\ 
    TeTriRF~\cite{wu2024tetrirf} & 30.43  & 0.906 & 0.25 & 0.062  & 14.9 \\ 
    \hline
    3DGStream~\cite{sun20243dgstream} & 31.58  & 0.941 & 0.14 & 7.8  & 1872 \\ 
    SpaceTimeGS~\cite{li2024spacetime}  & 32.05  & 0.946 & - & 6.7  & 1608 \\
    
    QUEEN~\cite{girish2024queen} & 32.19  & 0.946 & 0.14 & 0.75  & 180 \\
    
    GIFStream (high)~\cite{li2025gifstream} & 31.75  & 0.938 & 0.15 & 0.026  & 8.1 \\ 
    GIFStream (low)~\cite{li2025gifstream} & 30.22  & 0.934 & 0.17 & 0.016  & 3.9 \\ 
    \hline
    \textbf{\ours{}-AV1 (high)}  & 31.37  & 0.919 & 0.22 & 0.016  & 3.8 \\
    \textbf{\ours{}-AV1 (medium)}  & 31.00  & 0.910 & 0.24 & 0.012  & 2.9 \\ 
    \textbf{\ours{}-AV1 (low)}  & 29.55  & 0.870 & 0.34 & 0.003  & 0.7 \\ 

    \hline

    \textbf{\ours{}-DCVC (high)}  & 31.18  & 0.919 & 0.24 & 0.013  & 3.1 \\ 
    \textbf{\ours{}-DCVC (medium)}  & 30.63  & 0.904 & 0.28 & 0.008  & 1.9 \\
    \textbf{\ours{}-DCVC (low)}  & 29.81  & 0.890 & 0.31 & 0.005  & 1.2 \\

    \bottomrule
    \end{tabular}
    }
    \caption{
    Quantitative results on the N3D Video dataset. "Size" denotes the average client-side payload per frame, and "Bitrate" is computed at 30 FPS.
    }
    \label{tab:results_dynamic}
\end{table}

\noindent\textbf{Datasets.}
Previous work tends to focus exclusively on either static or dynamic benchmark, while our SCL can be evaluated across both settings by integrating with static/dynamic plane-factorized NeRF backbones~\cite{chen2022tensorf, wu2024tetrirf}. For static benchmarks, we use NeRF Synthetic~\cite{mildenhall2021nerf} and Tanks and Temples~\cite{liu2020neural} with the standard train/test splits. For dynamic benchmarks, we use Neural 3D Video (N3D)~\cite{li2022neural} and NHR~\cite{wu2020multi}, following the same evaluation protocol as prior dynamic NeRF compression work~\cite{wu2024tetrirf}.

\noindent\textbf{Baselines.}
We compare \ours{} against four groups of baselines: 
(1) parameter-efficient radiance fields that reduce model size without explicit RD optimization, such as Instant-NGP, Plenoxels, and BiRF~\cite{chen2022tensorf, muller2022instant, fridovich2022plenoxels, sun2022direct, fridovich2023k, shin2023binary};
(2) learned-codec-in-the-loop methods, such as CNC, NeRFCodec, NeRFPlayer, and VRVVC~\cite{chen2024far, lee2024ecrf, li2024nerfcodec, song2023nerfplayer, hu2025vrvvc};
(3) codec-agnostic methods that compress radiance fields with off-the-shelf codecs after training, including TeTriRF and PPNG~\cite{wu2024tetrirf, lee2025plenoptic}; and
(4) compressed 3DGS baselines, including HAC++~\cite{chen2025hac++} for static scenes and GIFStream~\cite{li2025gifstream} for dynamic scenes.


\noindent\textbf{Metrics.}
In addition to PSNR/SSIM/LPIPS versus model sizes, we additionally use BD-rate~\cite{bjontegaard2001calculation} to quantify rate–distortion (RD) performance. BD-rate measures the average bitrate difference between two RD curves at the same reconstruction quality. A lower BD-rate indicates fewer bits are required to achieve comparable quality. When reporting model size and overall bitrate, we count any parameters required for decoding 3D models at a client, including the learned, scene-specific codec/entropy-model parameters. All sizes are measured after quantization and entropy coding. Details on how we encode these parameters are presented in \cref{sec:baseline_payload} of the supplementary. 

\noindent\textbf{Implementation details.}
To ensure fair comparison across different TriPlane quality-size trade-offs, we build \ours{} directly on top of each baseline's released implementation. For static scenes, we use TensoRF~\cite{chen2022tensorf} as the backbone, matching NeRFCodec~\cite{li2024nerfcodec}. For dynamic scenes, we use the spatio-temporal DVGO backbone from TeTriRF~\cite{sun2022direct, wu2024tetrirf}. Following prior work~\cite{li2024nerfcodec}, we adopt a two-stage pipeline: we first train the vanilla backbone for each scene or a video segment to convergence, and finetune the radiance fields with SCL under a fixed codec configuration. We use a single NVIDIA V100 for both training and evaluation. Because STE treats the codec round trip as a forward-only operation, encoding and decoding need not be part of the GPU computation graph. For \ours{}-HEVC, we use PyNvVideoCodec, NVIDIA's Python-based library of Video Codec SDK, to measure the highest achievable hardware-accelerated decoding throughput for \cref{tab:decoding_speed}. For \ours{}-AV1, we employ PyAV built on FFmpeg in CPU since V100 does not provide hardware AV1 support.

\begin{table}[t!]
    \setlength{\tabcolsep}{4pt}  
    \centering
    \resizebox{\linewidth}{!}{
    \begin{tabular}{cccccc}
    \toprule
    & TensoRF~ & NeRFCodec & CNC & HAC++ &\textbf{\ours{}} \\
    \midrule
    Encoded feature planes & 69.95MB  & 0.05MB  & - & -&\textbf{0.34MB}  \\
    Encoded feature vectors & 0.05MB  & 0.05MB  & - & -&\textbf{0.05MB}  \\
    Encoded hash embeddings & - & - & 0.58MB  & -& - \\
    Neural codec parameters & - & 1.56MB  & - & -& - \\
    Entropy model parameters & - & -  & 0.01MB  & -& - \\
    Encoded anchor features & - & -  & - & 1.15MB & - \\
    MLPs (decoding)  & -  & -  & - & 0.21MB & -  \\
    MLPs (rendering)  & 0.04MB  & 0.03MB  & 0.18MB & - & \textbf{0.03MB}  \\
    \hline
    Total & 70.04MB & 1.69MB & 0.77MB & 1.36MB & \textbf{0.42MB} \\
    \bottomrule
    \end{tabular}
    }
    \caption{
    Memory breakdown on the NeRF Synthetic dataset. Learned-codec compression methods incur substantial client-side overhead because their scene-specific decoding parameters must be delivered together with the content. 3DGS compression methods require high-capacity features to represent the multiple attributes associated with each anchor point. In contrast, \ours{} maintains a compact volumetric representation while eliminating the need to transmit any scene-specific decoding parameters.
    }
    \label{tab:memory_breakdown}
\end{table}

\noindent\textbf{Static Benchmarks.} 
For static scenes, we pair \ours{} with JPEG to form \ours{}-JPEG, where rate control is achieved by varying the JPEG quality parameter. As shown in \cref{fig:catrf_image_quantitative}, \ours{}-JPEG forms a strong Pareto frontier against prior NeRF compression baselines. On NeRF Synthetic, it achieves an average 23\% bitrate reduction over CNC~\cite{chen2024far} at matched PSNR, confirming that directly adapting features to a standard codec can remain competitive with specialized learned compressors. Compared with HAC++~\cite{chen2025hac++}, \ours{} requires only 0.42 MB total client-side payload, whereas HAC++ requires 1.36 MB at similar quality. The detailed memory breakdown is presented in \cref{tab:memory_breakdown}, and the per-scene RD trade-off can be found in \cref{tab:per_scene_nerf} of the supplementary.
On the more challenging Tanks and Temples dataset, \ours{}-JPEG shows even larger gains, attaining an average 43\% bitrate saving over CNC~\cite{chen2024far} at matched PSNR and outperforming other baselines over a broad range of operating points. 
In \cref{fig:catrf_image_qualitative}, we show that \ours{}-JPEG offers a flexible operating range by adapting to various JPEG quality. It can operate at much lower bitrate with only modest loss of detail, or spend more bits to recover sharper edges and fewer artifacts than LCL and 3DGS methods at similar or lower rates. In contrast, previous pipelines that employ standard codecs are sensitive to codec artifacts, and therefore often need to operate at high-bitrate or even lossless settings, as exemplified by PPNG~\cite{lee2025plenoptic}.

\begin{table}[t!]
    \centering
    \resizebox{\linewidth}{!}{
    \begin{tabular}{cccc|ccc}
    \toprule
    Packing & Quantization & M & $\lambda_{tv}$ & PSNR$_\uparrow$ & LPIPS$_\downarrow$ & Mbps$_\downarrow$ \\
    \midrule
    FlattenGray & AbsMax&  128  & - & 31.19 & 0.272  & 3.15 \\ 
    FlattenGray & Channelwise&  128 & - & 31.47 & 0.260  & 14.29\\ 
    FlattenRGB & AbsMax&  128  & - & 30.92 & 0.289  & 2.41 \\ 
    FlattenRGB & Channelwise&  128 & - & 31.38 & 0.262 & 9.82 \\ 
    PixelShuffle & AbsMax&  128 & - & 30.67 & 0.313 & 2.82 \\ 
    PixelShuffle & Channelwise&  128 & - & 31.39 & 0.265 & 15.79\\  

    \hline
    FlattenGray & AbsMax&  128  & 5e-6 & 31.23 & 0.275  & 2.91 \\ 
    FlattenGray & AbsMax&  128  & 5e-4 & 30.41 & 0.316  & 1.59 \\ 
    
    \hline
    FlattenGray & AbsMax&  128  & 5e-5 & 30.75 & 0.298  & 1.83 \\ 
    FlattenRGB & AbsMax&  128  & 5e-5 & 30.43 & 0.312  & 1.69 \\ 
    PixelShuffle & AbsMax&  128 & 5e-5 & 30.08 & 0.341 & 2.11 \\ 

    \bottomrule
    \end{tabular}
    }
    \caption{Ablation results. Our main experiments adopt FlattenGray with AbsMax quantization as the default configuration. The results are obtained using \ours{}-AV1.}
    \label{tab:ablation}
\end{table}

\noindent\textbf{Dynamic Benchmarks.}
We pair both TeTriRF~\cite{wu2024tetrirf} and \ours{} with off-the-shelf video codecs (VP9/HEVC/AV1) and report RD curves together with BD-rate improvements. We also evaluate the state-of-the-art dynamic 3DGS compression method, GIFStream~\cite{li2025gifstream}, on the N3D dataset. Our framework can also integrate learned video codecs by freezing codec parameters during SCL finetuning. When we replace the standard codec with DCVC-RT~\cite{jia2025towards}, \ours{}-DCVC further improves the RD trade-off over \ours{}-AV1 at comparable quality levels, pushing the frontier to lower bitrates (see Fig.~\ref{fig:n3dv_rd_curve}). However, this configuration reintroduces a neural decoder on the client side, making deployment less lightweight than the purely standard-codec setting. On N3D, \ours{} consistently outperforms TeTriRF across all three standard codecs, achieving BD-rate reductions of 40.5\% (AV1), 63.3\% (HEVC), and 51.8\% (VP9). Compared with GIFStream~\cite{li2025gifstream}, \ours{}-AV1 reaches matched quality at substantially lower bitrate. The quantitative results are presented in \cref{tab:results_dynamic} and \cref{fig:n3dv_rd_curve}.  On NHR, \ours{} again strictly dominates TeTriRF, achieving higher quality at all tested bitrates and substantial bitrate savings. These results show that the SCL framework remains effective for dynamic scenes as well.



\noindent\textbf{Runtime Efficiency.}
The SCL framework does not modify the rendering pipeline of the underlying radiance-field backbone, so the rendering FPS of \ours{} is identical to that of the chosen backbone (e.g., TensoRF~\cite{chen2022tensorf} and DVGO~\cite{sun2022direct}). End-to-end client-side throughput, however, also depends on reconstruction from the received bitstreams. As shown in \cref{tab:decoding_speed}, the client-side decoding stage is the main bottleneck for compressed 3DGS methods: although GIFStream and HAC++ enjoy superior rendering FPS once reconstructed, they require MLP inference to recover Gaussian attributes. In contrast, the radiance fields reconstruction of \ours{} is neural-free, relying on standard-codec decoding only. This enables \ours{}-HEVC to reconstruct radiance fields at 138 FPS, versus 2.94 FPS/0.35 FPS for GIFStream/HAC++. As a result, \ours{} achieves the highest end-to-end throughput despite the limited rendering speed of NeRF backbones.

\begin{table}[t!]
    \centering
    \resizebox{\linewidth}{!}{
    \begin{tabular}{ccc|cccc}
    \toprule
    Packing & Quantization & M & PSNR$_\uparrow$ & LPIPS$_\downarrow$ & Mbps$_\downarrow$ & Training\\
    \midrule
    FlattenGray & AbsMax&  16  &  31.45 & 0.273  & 3.33 & 11h 57m\\ 
    FlattenGray & AbsMax&  32  &  31.13 & 0.285  & 2.37& 7h 53m\\ 
    FlattenGray & AbsMax&  64  & 30.75 & 0.298  & 1.92 & 5h 25m\\ 
    FlattenGray & AbsMax&  128  & 30.71 & 0.297  & 1.83 & 2h 33m\\ 
    FlattenGray & AbsMax&  256  & 30.67 & 0.302  & 1.84 & 1h 53m\\ 
    \bottomrule
    \end{tabular}
    }
    \caption{
    Ablation on the caching refresh interval $M$. Larger $M$ substantially reduces training time and bitrate with only a modest quality degradation.}
    \label{tab:ablation_cache}
\end{table}

\begin{table}[t!]
    \setlength{\tabcolsep}{4pt}  
    \centering
    \resizebox{\linewidth}{!}{
    \begin{tabular}{ccccc}
    \toprule
    & MLP for Recon. & Decoding & Rendering & End-to-end \\
    \midrule
    NeRFCodec~\cite{li2024nerfcodec} & Yes & 0.08 FPS & 1.56 FPS & 0.07 FPS \\  
    CNC~\cite{chen2024far} & No & 0.98 FPS & 5.16 FPS & 0.82 FPS \\
    HAC++~\cite{chen2025hac++} & Yes & 0.35 FPS & 169 FPS & 0.35 FPS \\
    GIFStream~\cite{li2025gifstream} & Yes & 2.94 FPS & 268 FPS & 2.91 FPS \\ 
    \ours{}-HEVC & No & 138 FPS & 5.86 FPS & 5.62 FPS \\ 
    \bottomrule
    \end{tabular}
    }
    \caption{
    Decoding and rendering speed comparison. We report decoding FPS, rendering FPS, and end-to-end throughput including both stages. \ours{} does not require any ML inference to retrieve 3D representations from the encoded bitstreams.
    }
    \label{tab:decoding_speed}
\end{table}

\noindent\textbf{Ablation study.}
\cref{tab:ablation} ablates the main design choices of \ours{}, including feature-plane packing, quantization scheme, and total-variation (TV) regularization.
FlattenGray consistently yields the best RD trade-off, while FlattenRGB and PixelShuffle achieve similar or lower quality at only slightly lower bitrates.
This suggests that avoiding cross-channel mixing makes codec artifacts less harmful to radiance-field semantics.
Although channelwise quantization improves fidelity, it increases bitrate by roughly $4$-$5\times$, making it unsuitable for the streaming regime. We therefore use FlattenGray with AbsMax quantization in our main experiments.
\cref{tab:ablation} also varies the TV weight $\lambda_{tv}$. Moderate TV regularization improves codec compatibility without overly blurring high-frequency details.
In \cref{tab:ablation_cache}, we ablate the cache refresh interval $M$ under a fixed TV weight ($\lambda_{tv}=5\times10^{-5}$). We empirically observed that $M{=}128$ reduces training time and bitrate with only modest quality degradation, so we configured $M{=}128$ in the main experiments.

\section{Conclusion}
\label{sec:conclusion}

We presented \ours{}, a standard-codec-in-the-loop compression framework for plane-factorized radiance fields. By quantizing and packing feature planes into codec-friendly canvases, running a real JPEG/VP9/HEVC/AV1 round trip, and propagating gradients with a straight-through estimator, \ours{} directly optimizes the decoded radiance fields that clients actually render, without requiring any per-scene neural codec or entropy-model parameters. Across static and dynamic benchmarks, \ours{} consistently improves the rate--distortion trade-off over codec-agnostic and learned-codec-in-the-loop baselines, achieving up to 63.3\% BD-rate reduction. It also outperforms state-of-the-art compressed 3DGS methods in both compression efficiency and decoding speed, highlighting a practical path toward low-bitrate, compression-resilient volumetric media streaming.



\section*{Acknowledgments}
\label{sec:ack}
\small
This work was supported in part by the National Science Foundation (NSF) under Grants CNS~2106463 and CNS~1901137.
{
    \small
    \bibliographystyle{ieeenat_fullname}
    \bibliography{main}

@String(TOG= {ACM Trans. Graph.})

@String(ICASSP=	{ICASSP})

@String(ICIP = {IEEE Int. Conf. Image Process.})

@String(ICLR = {Int. Conf. Learn. Represent.})

@String(AAAI = {AAAI})

@String(TOG   = {ACM TOG})

@String(ICIP  = {ICIP})

@String(ICLR  = {ICLR})

@article{muller2022instant,
  title={Instant neural graphics primitives with a multiresolution hash encoding},
  author={M{\"u}ller, Thomas and Evans, Alex and Schied, Christoph and Keller, Alexander},
  journal={ACM transactions on graphics (TOG)},
  volume={41},
  number={4},
  pages={1--15},
  year={2022},
  publisher={ACM New York, NY, USA}
}

@article{shin2023binary,
  title={Binary radiance fields},
  author={Shin, Seungjoo and Park, Jaesik},
  journal={Advances in neural information processing systems},
  volume={36},
  pages={55919--55931},
  year={2023}
}

@inproceedings{fridovich2022plenoxels,
  title={Plenoxels: Radiance fields without neural networks},
  author={Fridovich-Keil, Sara and Yu, Alex and Tancik, Matthew and Chen, Qinhong and Recht, Benjamin and Kanazawa, Angjoo},
  booktitle={Proceedings of the IEEE/CVF conference on computer vision and pattern recognition},
  pages={5501--5510},
  year={2022}
}

@inproceedings{chan2022efficient,
  title={Efficient geometry-aware 3d generative adversarial networks},
  author={Chan, Eric R and Lin, Connor Z and Chan, Matthew A and Nagano, Koki and Pan, Boxiao and De Mello, Shalini and Gallo, Orazio and Guibas, Leonidas J and Tremblay, Jonathan and Khamis, Sameh and others},
  booktitle={Proceedings of the IEEE/CVF conference on computer vision and pattern recognition},
  pages={16123--16133},
  year={2022}
}

@inproceedings{fridovich2023k,
  title={K-planes: Explicit radiance fields in space, time, and appearance},
  author={Fridovich-Keil, Sara and Meanti, Giacomo and Warburg, Frederik Rahb{\ae}k and Recht, Benjamin and Kanazawa, Angjoo},
  booktitle={Proceedings of the IEEE/CVF Conference on Computer Vision and Pattern Recognition},
  pages={12479--12488},
  year={2023}
}

@inproceedings{chen2022tensorf,
  title={Tensorf: Tensorial radiance fields},
  author={Chen, Anpei and Xu, Zexiang and Geiger, Andreas and Yu, Jingyi and Su, Hao},
  booktitle={European conference on computer vision},
  pages={333--350},
  year={2022},
  organization={Springer}
}

@inproceedings{sun2022direct,
  title={Direct voxel grid optimization: Super-fast convergence for radiance fields reconstruction},
  author={Sun, Cheng and Sun, Min and Chen, Hwann-Tzong},
  booktitle={Proceedings of the IEEE/CVF conference on computer vision and pattern recognition},
  pages={5459--5469},
  year={2022}
}

@article{li2022streaming,
  title={Streaming radiance fields for 3d video synthesis},
  author={Li, Lingzhi and Shen, Zhen and Wang, Zhongshu and Shen, Li and Tan, Ping},
  journal={Advances in Neural Information Processing Systems},
  volume={35},
  pages={13485--13498},
  year={2022}
}

@article{song2023nerfplayer,
  title={Nerfplayer: A streamable dynamic scene representation with decomposed neural radiance fields},
  author={Song, Liangchen and Chen, Anpei and Li, Zhong and Chen, Zhang and Chen, Lele and Yuan, Junsong and Xu, Yi and Geiger, Andreas},
  journal={IEEE Transactions on Visualization and Computer Graphics},
  volume={29},
  number={5},
  pages={2732--2742},
  year={2023},
  publisher={IEEE}
}

@inproceedings{zheng2024jointrf,
  title={JointRF: end-to-end joint optimization for dynamic neural radiance field representation and compression},
  author={Zheng, Zihan and Zhong, Houqiang and Hu, Qiang and Zhang, Xiaoyun and Song, Li and Zhang, Ya and Wang, Yanfeng},
  booktitle={2024 IEEE International Conference on Image Processing (ICIP)},
  pages={3292--3298},
  year={2024},
  organization={IEEE}
}

@inproceedings{zhang2024rate,
  title={Rate-aware compression for nerf-based volumetric video},
  author={Zhang, Zhiyu and Lu, Guo and Liang, Huanxiong and Cheng, Zhengxue and Tang, Anni and Song, Li},
  booktitle={Proceedings of the 32nd ACM International Conference on Multimedia},
  pages={3974--3983},
  year={2024}
}

@inproceedings{hu2025vrvvc,
  title={VRVVC: Variable-Rate NeRF-Based Volumetric Video Compression},
  author={Hu, Qiang and Zhong, Houqiang and Zheng, Zihan and Zhang, Xiaoyun and Cheng, Zhengxue and Song, Li and Zhai, Guangtao and Wang, Yanfeng},
  booktitle={Proceedings of the AAAI Conference on Artificial Intelligence},
  volume={39},
  number={4},
  pages={3563--3571},
  year={2025}
}

@inproceedings{chen2024far,
  title={How far can we compress instant-ngp-based nerf?},
  author={Chen, Yihang and Wu, Qianyi and Harandi, Mehrtash and Cai, Jianfei},
  booktitle={Proceedings of the IEEE/CVF Conference on Computer Vision and Pattern Recognition},
  pages={20321--20330},
  year={2024}
}

@inproceedings{lee2024ecrf,
  title={Ecrf: Entropy-constrained neural radiance fields compression with frequency domain optimization},
  author={Lee, Soonbin and Shu, Fangwen and Sanchez, Yago and Schierl, Thomas and Hellge, Cornelius},
  booktitle={2024 IEEE 26th International Workshop on Multimedia Signal Processing (MMSP)},
  pages={1--6},
  year={2024},
  organization={IEEE}
}

@inproceedings{wu2024tetrirf,
  title={Tetrirf: Temporal tri-plane radiance fields for efficient free-viewpoint video},
  author={Wu, Minye and Wang, Zehao and Kouros, Georgios and Tuytelaars, Tinne},
  booktitle={Proceedings of the IEEE/CVF conference on computer vision and pattern recognition},
  pages={6487--6496},
  year={2024}
}

@inproceedings{li2024nerfcodec,
  title={Nerfcodec: Neural feature compression meets neural radiance fields for memory-efficient scene representation},
  author={Li, Sicheng and Li, Hao and Liao, Yiyi and Yu, Lu},
  booktitle={Proceedings of the IEEE/CVF Conference on Computer Vision and Pattern Recognition},
  pages={21274--21283},
  year={2024}
}

@inproceedings{lee2025plenoptic,
  title={Plenoptic png: Real-time neural radiance fields in 150 kb},
  author={Lee, Jae Yong and Wu, Yuqun and Zou, Chuhang and Hoiem, Derek and Wang, Shenlong},
  booktitle={2025 International Conference on 3D Vision (3DV)},
  pages={502--511},
  year={2025},
  organization={IEEE}
}

@inproceedings{kang2025codecnerf,
  title={Codecnerf: Toward fast encoding and decoding, compact, and high-quality novel-view synthesis},
  author={Kang, Gyeongjin and Lee, Younggeun and Oh, Seungjun and Park, Eunbyung},
  booktitle={Proceedings of the AAAI Conference on Artificial Intelligence},
  volume={39},
  number={4},
  pages={4203--4211},
  year={2025}
}

@inproceedings{chen2024hac,
  title={Hac: Hash-grid assisted context for 3d gaussian splatting compression},
  author={Chen, Yihang and Wu, Qianyi and Lin, Weiyao and Harandi, Mehrtash and Cai, Jianfei},
  booktitle={European Conference on Computer Vision},
  pages={422--438},
  year={2024},
  organization={Springer}
}

@inproceedings{niedermayr2024compressed,
  title={Compressed 3d gaussian splatting for accelerated novel view synthesis},
  author={Niedermayr, Simon and Stumpfegger, Josef and Westermann, R{\"u}diger},
  booktitle={Proceedings of the IEEE/CVF Conference on Computer Vision and Pattern Recognition},
  pages={10349--10358},
  year={2024}
}

@inproceedings{wang2024videorf,
  title={Videorf: Rendering dynamic radiance fields as 2d feature video streams},
  author={Wang, Liao and Yao, Kaixin and Guo, Chengcheng and Zhang, Zhirui and Hu, Qiang and Yu, Jingyi and Xu, Lan and Wu, Minye},
  booktitle={Proceedings of the IEEE/CVF Conference on Computer Vision and Pattern Recognition},
  pages={470--481},
  year={2024}
}

@inproceedings{li2025gifstream,
  title={Gifstream: 4d gaussian-based immersive video with feature stream},
  author={Li, Hao and Li, Sicheng and Gao, Xiang and Batuer, Abudouaihati and Yu, Lu and Liao, Yiyi},
  booktitle={Proceedings of the Computer Vision and Pattern Recognition Conference},
  pages={21761--21770},
  year={2025}
}

@article{chen2025hac++,
  title={Hac++: Towards 100x compression of 3d gaussian splatting},
  author={Chen, Yihang and Wu, Qianyi and Lin, Weiyao and Harandi, Mehrtash and Cai, Jianfei},
  journal={IEEE Transactions on Pattern Analysis and Machine Intelligence},
  year={2025},
  publisher={IEEE}
}

@inproceedings{li2024spacetime,
  title={Spacetime gaussian feature splatting for real-time dynamic view synthesis},
  author={Li, Zhan and Chen, Zhang and Li, Zhong and Xu, Yi},
  booktitle={Proceedings of the IEEE/CVF Conference on Computer Vision and Pattern Recognition},
  pages={8508--8520},
  year={2024}
}

@inproceedings{wu20244d,
  title={4d gaussian splatting for real-time dynamic scene rendering},
  author={Wu, Guanjun and Yi, Taoran and Fang, Jiemin and Xie, Lingxi and Zhang, Xiaopeng and Wei, Wei and Liu, Wenyu and Tian, Qi and Wang, Xinggang},
  booktitle={Proceedings of the IEEE/CVF conference on computer vision and pattern recognition},
  pages={20310--20320},
  year={2024}
}

@inproceedings{shaw2024swings,
  title={Swings: sliding windows for dynamic 3d gaussian splatting},
  author={Shaw, Richard and Nazarczuk, Michal and Song, Jifei and Moreau, Arthur and Catley-Chandar, Sibi and Dhamo, Helisa and P{\'e}rez-Pellitero, Eduardo},
  booktitle={European Conference on Computer Vision},
  pages={37--54},
  year={2024},
  organization={Springer}
}

@article{girish2024queen,
  title={Queen: Quantized efficient encoding of dynamic gaussians for streaming free-viewpoint videos},
  author={Girish, Sharath and Li, Tianye and Mazumdar, Amrita and Shrivastava, Abhinav and De Mello, Shalini and others},
  journal={Advances in Neural Information Processing Systems},
  volume={37},
  pages={43435--43467},
  year={2024}
}

@inproceedings{sun20243dgstream,
  title={3dgstream: On-the-fly training of 3d gaussians for efficient streaming of photo-realistic free-viewpoint videos},
  author={Sun, Jiakai and Jiao, Han and Li, Guangyuan and Zhang, Zhanjie and Zhao, Lei and Xing, Wei},
  booktitle={Proceedings of the IEEE/CVF Conference on Computer Vision and Pattern Recognition},
  pages={20675--20685},
  year={2024}
}

@article{lee2025compression,
  title={Compression of 3d gaussian splatting with optimized feature planes and standard video codecs},
  author={Lee, Soonbin and Shu, Fangwen and Sanchez, Yago and Schierl, Thomas and Hellge, Cornelius},
  journal={arXiv preprint arXiv:2501.03399},
  year={2025}
}

@article{wang2024v,
  title={V\^{} 3: Viewing Volumetric Videos on Mobiles via Streamable 2D Dynamic Gaussians},
  author={Wang, Penghao and Zhang, Zhirui and Wang, Liao and Yao, Kaixin and Xie, Siyuan and Yu, Jingyi and Wu, Minye and Xu, Lan},
  journal={ACM Transactions on Graphics (TOG)},
  volume={43},
  number={6},
  pages={1--13},
  year={2024},
  publisher={ACM New York, NY, USA}
}

@inproceedings{li2022neural,
  title={Neural 3d video synthesis from multi-view video},
  author={Li, Tianye and Slavcheva, Mira and Zollhoefer, Michael and Green, Simon and Lassner, Christoph and Kim, Changil and Schmidt, Tanner and Lovegrove, Steven and Goesele, Michael and Newcombe, Richard and others},
  booktitle={Proceedings of the IEEE/CVF conference on computer vision and pattern recognition},
  pages={5521--5531},
  year={2022}
}

@article{mildenhall2021nerf,
  title={Nerf: Representing scenes as neural radiance fields for view synthesis},
  author={Mildenhall, Ben and Srinivasan, Pratul P and Tancik, Matthew and Barron, Jonathan T and Ramamoorthi, Ravi and Ng, Ren},
  journal={Communications of the ACM},
  volume={65},
  number={1},
  pages={99--106},
  year={2021},
  publisher={ACM New York, NY, USA}
}

@article{liu2020neural,
  title={Neural sparse voxel fields},
  author={Liu, Lingjie and Gu, Jiatao and Zaw Lin, Kyaw and Chua, Tat-Seng and Theobalt, Christian},
  journal={Advances in Neural Information Processing Systems},
  volume={33},
  pages={15651--15663},
  year={2020}
}

@inproceedings{wu2020multi,
  title={Multi-view neural human rendering},
  author={Wu, Minye and Wang, Yuehao and Hu, Qiang and Yu, Jingyi},
  booktitle={Proceedings of the IEEE/CVF Conference on Computer Vision and Pattern Recognition},
  pages={1682--1691},
  year={2020}
}

@article{jin2023capture,
  title={From capture to display: A survey on volumetric video},
  author={Jin, Yili and Hu, Kaiyuan and Liu, Junhua and Wang, Fangxin and Liu, Xue},
  journal={arXiv preprint arXiv:2309.05658},
  year={2023}
}

@inproceedings{orts2016holoportation,
  title={Holoportation: Virtual 3d teleportation in real-time},
  author={Orts-Escolano, Sergio and Rhemann, Christoph and Fanello, Sean and Chang, Wayne and Kowdle, Adarsh and Degtyarev, Yury and Kim, David and Davidson, Philip L and Khamis, Sameh and Dou, Mingsong and others},
  booktitle={Proceedings of the 29th annual symposium on user interface software and technology},
  pages={741--754},
  year={2016}
}

@article{collet2015high,
  title={High-quality streamable free-viewpoint video},
  author={Collet, Alvaro and Chuang, Ming and Sweeney, Pat and Gillett, Don and Evseev, Dennis and Calabrese, David and Hoppe, Hugues and Kirk, Adam and Sullivan, Steve},
  journal={ACM Transactions on Graphics (ToG)},
  volume={34},
  number={4},
  pages={1--13},
  year={2015},
  publisher={ACM New York, NY, USA}
}

@inproceedings{bentaleb2022low,
  title={Low latency live streaming implementation in dash and hls},
  author={Bentaleb, Abdelhak and Zhan, Zhengdao and Tashtarian, Farzad and Lim, May and Harous, Saad and Timmerer, Christian and Hellwagner, Hermann and Zimmermann, Roger},
  booktitle={Proceedings of the 30th ACM International Conference on Multimedia},
  pages={7343--7346},
  year={2022}
}

@article{graziosi2020overview,
  title={An overview of ongoing point cloud compression standardization activities: Video-based (V-PCC) and geometry-based (G-PCC)},
  author={Graziosi, Danillo and Nakagami, Ohji and Kuma, Satoru and Zaghetto, Alexandre and Suzuki, Teruhiko and Tabatabai, Ali},
  journal={APSIPA Transactions on Signal and Information Processing},
  volume={9},
  pages={e13},
  year={2020},
  publisher={Cambridge University Press}
}

@article{sodagar2011mpeg,
  title={The mpeg-dash standard for multimedia streaming over the internet},
  author={Sodagar, Iraj},
  journal={IEEE multimedia},
  volume={18},
  number={4},
  pages={62--67},
  year={2011},
  publisher={IEEE}
}

@article{sani2017adaptive,
  title={Adaptive bitrate selection: A survey},
  author={Sani, Yusuf and Mauthe, Andreas and Edwards, Christopher},
  journal={IEEE Communications Surveys \& Tutorials},
  volume={19},
  number={4},
  pages={2985--3014},
  year={2017},
  publisher={IEEE}
}

@article{bengio2013estimating,
  title={Estimating or propagating gradients through stochastic neurons for conditional computation},
  author={Bengio, Yoshua and L{\'e}onard, Nicholas and Courville, Aaron},
  journal={arXiv preprint arXiv:1308.3432},
  year={2013}
}

@article{yang2025improving,
  title={Improving the Straight-Through Estimator with Zeroth-Order Information},
  author={Yang, Ningfeng and Aamodt, Tor M},
  journal={arXiv preprint arXiv:2510.23926},
  year={2025}
}

@article{mack2025efficient,
  title={Efficient evaluation of quantization-effects in neural codecs},
  author={Mack, Wolfgang and Mustafa, Ahmed and {\L}aganowski, Rafa{\l} and Hijazy, Samer},
  journal={arXiv preprint arXiv:2502.04770},
  year={2025}
}

@article{spall1998overview,
  title={An overview of the simultaneous perturbation method for efficient optimization},
  author={Spall, James C},
  journal={Johns Hopkins apl technical digest},
  volume={19},
  number={4},
  pages={482--492},
  year={1998}
}

@inproceedings{ramirez2021differentiable,
  title={Differentiable signal processing with black-box audio effects},
  author={Ram{\'\i}rez, Marco A Mart{\'\i}nez and Wang, Oliver and Smaragdis, Paris and Bryan, Nicholas J},
  booktitle={ICASSP 2021-2021 IEEE International Conference on Acoustics, Speech and Signal Processing (ICASSP)},
  pages={66--70},
  year={2021},
  organization={IEEE}
}

@article{gupta20233dgen,
  title={3dgen: Triplane latent diffusion for textured mesh generation},
  author={Gupta, Anchit and Xiong, Wenhan and Nie, Yixin and Jones, Ian and O{\u{g}}uz, Barlas},
  journal={arXiv preprint arXiv:2303.05371},
  year={2023}
}

@inproceedings{shue20233d,
  title={3d neural field generation using triplane diffusion},
  author={Shue, J Ryan and Chan, Eric Ryan and Po, Ryan and Ankner, Zachary and Wu, Jiajun and Wetzstein, Gordon},
  booktitle={Proceedings of the IEEE/CVF Conference on Computer Vision and Pattern Recognition},
  pages={20875--20886},
  year={2023}
}

@article{li2024instant3d,
  title={Instant3d: Instant text-to-3d generation},
  author={Li, Ming and Zhou, Pan and Liu, Jia-Wei and Keppo, Jussi and Lin, Min and Yan, Shuicheng and Xu, Xiangyu},
  journal={International Journal of Computer Vision},
  volume={132},
  number={10},
  pages={4456--4472},
  year={2024},
  publisher={Springer}
}

@inproceedings{yeo2018neural,
  title={Neural adaptive content-aware internet video delivery},
  author={Yeo, Hyunho and Jung, Youngmok and Kim, Jaehong and Shin, Jinwoo and Han, Dongsu},
  booktitle={13th USENIX Symposium on Operating Systems Design and Implementation (OSDI 18)},
  pages={645--661},
  year={2018}
}

@inproceedings{yeo2020nemo,
  title={Nemo: enabling neural-enhanced video streaming on commodity mobile devices},
  author={Yeo, Hyunho and Chong, Chan Ju and Jung, Youngmok and Ye, Juncheol and Han, Dongsu},
  booktitle={Proceedings of the 26th Annual International Conference on Mobile Computing and Networking},
  pages={1--14},
  year={2020}
}

@inproceedings{isik2023sandwiched,
  title={Sandwiched video compression: Efficiently extending the reach of standard codecs with neural wrappers},
  author={Isik, Berivan and Guleryuz, Onur G and Tang, Danhang and Taylor, Jonathan and Chou, Philip A},
  booktitle={2023 IEEE International Conference on Image Processing (ICIP)},
  pages={2055--2059},
  year={2023},
  organization={IEEE}
}

@article{bentaleb2025solutions,
  title={Solutions, challenges, and opportunities in volumetric video streaming: an architectural perspective},
  author={Bentaleb, Abdelhak and Lim, May and Hammoudi, Sarra and Harous, Saad and Zimmermann, Roger},
  journal={ACM Transactions on Multimedia Computing, Communications and Applications},
  volume={21},
  number={7},
  pages={1--35},
  year={2025},
  publisher={ACM New York, NY}
}

@inproceedings{courbariaux2015binaryconnect,
  title={BinaryConnect: Training Deep Neural Networks with Binary Weights During Propagations},
  author={Courbariaux, Matthieu and Bengio, Yoshua and David, Jean-Pierre},
  booktitle={NeurIPS},
  year={2015}
}

@inproceedings{balle2017iclr,
  title={End-to-End Optimized Image Compression},
  author={Ballé, Johannes and Laparra, Valero and Simoncelli, Eero P.},
  booktitle={ICLR},
  year={2017}
}

@article{balle2018tisl,
  title={Variational Image Compression with a Scale Hyperprior},
  author={Ballé, Johannes and Minnen, David and Singh, Saurabh and Johnston, Sung Jin Hwang and Simoncelli, Eero P.},
  journal={IEEE Transactions on Image Processing},
  year={2018}
}

@inproceedings{cai2019proxylessnas,
  title={ProxylessNAS: Direct Neural Architecture Search on Target Task and Hardware},
  author={Cai, Han and Zhu, Ligeng and Han, Song},
  booktitle={ICLR},
  year={2019}
}

@inproceedings{li2024neural,
  title={Neural video compression with feature modulation},
  author={Li, Jiahao and Li, Bin and Lu, Yan},
  booktitle={Proceedings of the IEEE/CVF Conference on Computer Vision and Pattern Recognition},
  pages={26099--26108},
  year={2024}
}

@inproceedings{jia2025towards,
  title={Towards practical real-time neural video compression},
  author={Jia, Zhaoyang and Li, Bin and Li, Jiahao and Xie, Wenxuan and Qi, Linfeng and Li, Houqiang and Lu, Yan},
  booktitle={Proceedings of the Computer Vision and Pattern Recognition Conference},
  pages={12543--12552},
  year={2025}
}

@article{bjontegaard2001calculation,
  title={Calculation of average PSNR differences between RD-curves},
  author={Bjontegaard, Gisle},
  journal={ITU-T SG16, Doc. VCEG-M33},
  year={2001}
}
}

\clearpage
\setcounter{page}{1}
\maketitlesupplementary
\appendix

\section{Caching for SCL Training}
\label{sec:ste_cache}
If we naively invoke the codec round trip $\mathcal{C}_q$ on every mini-batch, the encoding cost quickly dominates training time, especially for dynamic scenes where a video codec must operate on a sequence of frames (see ~\cref{tab:ablation_cache}).
To make our standard-codec-in-the-loop (SCL) training practical, we maintain a per-frame cache of decoded feature planes and density volumes $(\widehat{\mathcal{P}}, \widehat{D})$, and only refresh this cache when necessary. For each frame $t$ and each axis $ax\!\in\!\{xy,xz,yz\}$ we cache the decoded TriPlane $\widehat{P}^{ax}_t$ and a snapshot of the raw plane $P^{ax}_t$ at the time of refresh. We do the same for the density grid $D_t$. Our refresh policy consists of two cases:

\noindent
\textbf{Step-based refresh.}
We keep track of the last refresh step $g_{\text{cache}}$ and the current global step $g$. Every $M$ training steps, i.e., when $g - g_{\text{cache}} \ge M$, we re-run the codec over the entire video segment to update the cache.

\noindent
\textbf{Change-based refresh.}
Step-based refresh alone may lag behind rapid parameter changes early in training. We therefore also monitor how much the current TriPlane has drifted from the last cached snapshot. If the relative change between the current planes and their cached snapshots exceeds a threshold $\epsilon$ (measured by a normalized $\ell_2$ difference), we trigger an early refresh even if $g - g_{\text{cache}} < M$. In practice, this makes the cache more responsive at the beginning of training while keeping codec round trips infrequent once the model stabilizes.

Putting these pieces together, the SCL training loop with caching can be summarized as:

\begin{enumerate}
\item Sample a frame index $t$, rays $\{(\mathbf{o},\mathbf{d})\}$, and camera pose $\pi$.

\item If the cache is empty or $g - g_{\text{cache}} \ge M$, mark \emph{refresh} as true.
\item If the relative change between the current $(P^{ax}_t, D_t)$ and their cached snapshots exceeds a threshold $\epsilon$, also mark \emph{refresh} as true.
\item If \emph{refresh} is true, then run the encode–decode codec round trip (as illustrated in Fig.~\ref{fig:eval_workflow}) for the video segment.
\item Update $g_{\text{cache}} \leftarrow g$.

\begin{figure}[t]
  \centering

   \includegraphics[width=1.0\linewidth]{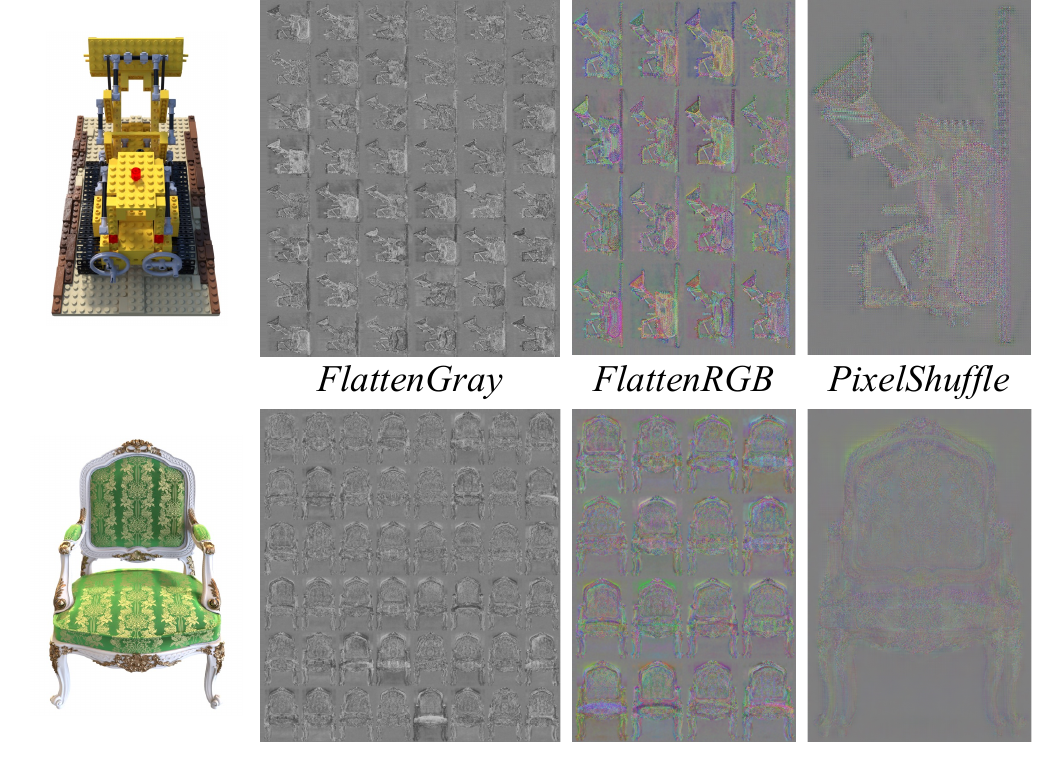}

   \caption{
    Visualization of appearance-plane canvases under different packing strategies. All three layouts represent the same TriPlane features, but they induce different spatial statistics and texture patterns, which affect how well standard codecs exploit redundancy and preserve feature semantics after compression.
    }
   \label{fig:packing_example}
\end{figure}

\item STE substitution using cached reconstructions: 
  \begin{enumerate}
  \item For each axis $ax\in\{xy,xz,yz\}$:
    \[
      \widehat{P}^{ax}_t \leftarrow \text{cached decoded plane},
    \]
    \[
      \widetilde{P}^{ax}_t \leftarrow \widehat{P}^{ax}_t + \big(P^{ax}_t - \texttt{detach}(P^{ax}_t)\big).
    \]
  \item For density:
    \[
      \widehat{D}_t \leftarrow \text{cached decoded density}, 
    \]
    \[
      \widetilde{D}_t \leftarrow \widehat{D}_t + \big(D_t - \texttt{detach}(D_t)\big).
    \]
  \end{enumerate}

\item Render and compute losses:  
\[
I\leftarrow \mathcal{R}(\widetilde{\mathcal{P}}_t,\widetilde{D}_t,\pi;\phi),
\]
\end{enumerate}

\cref{tab:ablation_cache} shows that increasing the refresh interval $M$ yields a favorable trade-off between accuracy and training efficiency. It suggests that relatively infrequent cache updates (e.g., $M = 128$) already capture most of the benefit of SCL training, while keeping the overhead of expensive codec round trips manageable. 

%


\section{Decoder-Side Payload of Baseline Methods}
\label{sec:baseline_payload}

In a streaming scenario, NeRF compression methods that train a \emph{scene-specific} codec (e.g., NeRFCodec~\cite{li2024nerfcodec}, Rate-aware NeRF~\cite{zhang2024rate}, and CNC~\cite{chen2024far}) must transmit not only the encoded radiance-field representation but also the decoder-side codec parameters required to reconstruct NeRFs at the client. While prior work typically reports only the feature bitstream or backbone model size, in our evaluation these decoder parameters are part of the content payload and therefore must be included in the bitrate budget. In contrast, \ours{} is encoded and decoded using standard codecs that are ubiquitous on commodity devices, avoiding this additional overhead (see Tab.~\ref{tab:memory_breakdown} in the main paper).

\paragraph{Observed payload for NeRFCodec.}
For NeRFCodec~\cite{li2024nerfcodec}, we adopt their official implementation, which uses \texttt{Cheng2020Anchor} neural image codec from CompressAI. Empirically, we observed that the following decoder-side modules are fine-tuned together with the radiance-field backbone on a per-scene basis, and thus must be transmitted alongside the encoded NeRFs for decoding:

\begin{itemize}[leftmargin=1.2em,topsep=2pt,itemsep=2pt]
  \item \texttt{decoder\_adaptor}
  \item \texttt{context\_prediction}
  \item \texttt{entropy\_parameters}
  \item \texttt{entropy\_bottleneck.quantiles}
\end{itemize}


Note that NeRFCodec employs two separate neural image codecs for the appearance and density streams, respectively. Encoder-side modules, although also fine-tuned, are excluded from the bitrate budget since they are not required at the client for decoding.
The authors of NeRFCodec do not release their bitrate budget estimation code. Rather than reporting the raw parameter size, we apply per-tensor quantization and entropy coding to estimate a more realistic compressed size:

\begin{enumerate}[leftmargin=1.5em,topsep=2pt,itemsep=2pt]
  \item \textbf{Per-tensor quantization.} For each tensor, we collect its values $x \in \mathbb{R}$ and compute
  \[
    x_{\min} = \min(x), \quad x_{\max} = \max(x).
  \]
  We quantize to $q_{\text{bits}}$ bits (we use $q_{\text{bits}} = 8$ in our experiments). Let $L = 2^{q_{\text{bits}}}$ and
  \[
    \mathrm{scale} = \frac{x_{\max} - x_{\min}}{L - 1}.
  \]
  Each element is mapped to an integer level
  \[
    q = \mathrm{round}\!\left(\frac{x - x_{\min}}{\mathrm{scale}}\right),
  \]
  and clamped to $[0, L-1]$.
  \item \textbf{Entropy estimate.} We compute a histogram over the quantized levels and derive the empirical PMF $p_i$ over $\{0,\dots,L-1\}$. The Shannon bound gives an idealized bit cost
  \[
    H_{\text{bits}} \;=\; N \sum_{i} -p_i \log_2 p_i,
  \]
  where $N$ is the number of elements and the sum is taken over non-zero probabilities.
  \item \textbf{Per-tensor header.} To reconstruct a tensor from its quantized form, the decoder also needs lightweight side information: $(x_{\min}, x_{\max})$, $q_{\text{bits}}$, and the tensor shape. We therefore add a small header
  \[
    \mathrm{header\_bits}
    = 2 \cdot 32 \;+\; 8 \;+\; 32 \cdot d,
  \]
  where $d$ is the number of dimensions.
\end{enumerate}
The total bit cost for a tensor is therefore
\[
  \mathrm{bits}
  = H_{\text{bits}} + \mathrm{header\_bits},
\]
which is how we estimate the bitrate budget required for NeRFCodec's decoder-side parameters.

The compression procedure described above yields a raw decoder payload of roughly $24$~MB in total (about $11.9$~MB for the density codec and $12.8$~MB for the appearance codec). After compression, this reduces to about $1.6$~MB. These quantities are what we report as NeRFCodec's memory footprint, which represents the actual payload in a streaming application if we use NeRFCodec as the 3D representation.

\begin{figure}[t]
  \centering
   \includegraphics[width=1.0\linewidth]{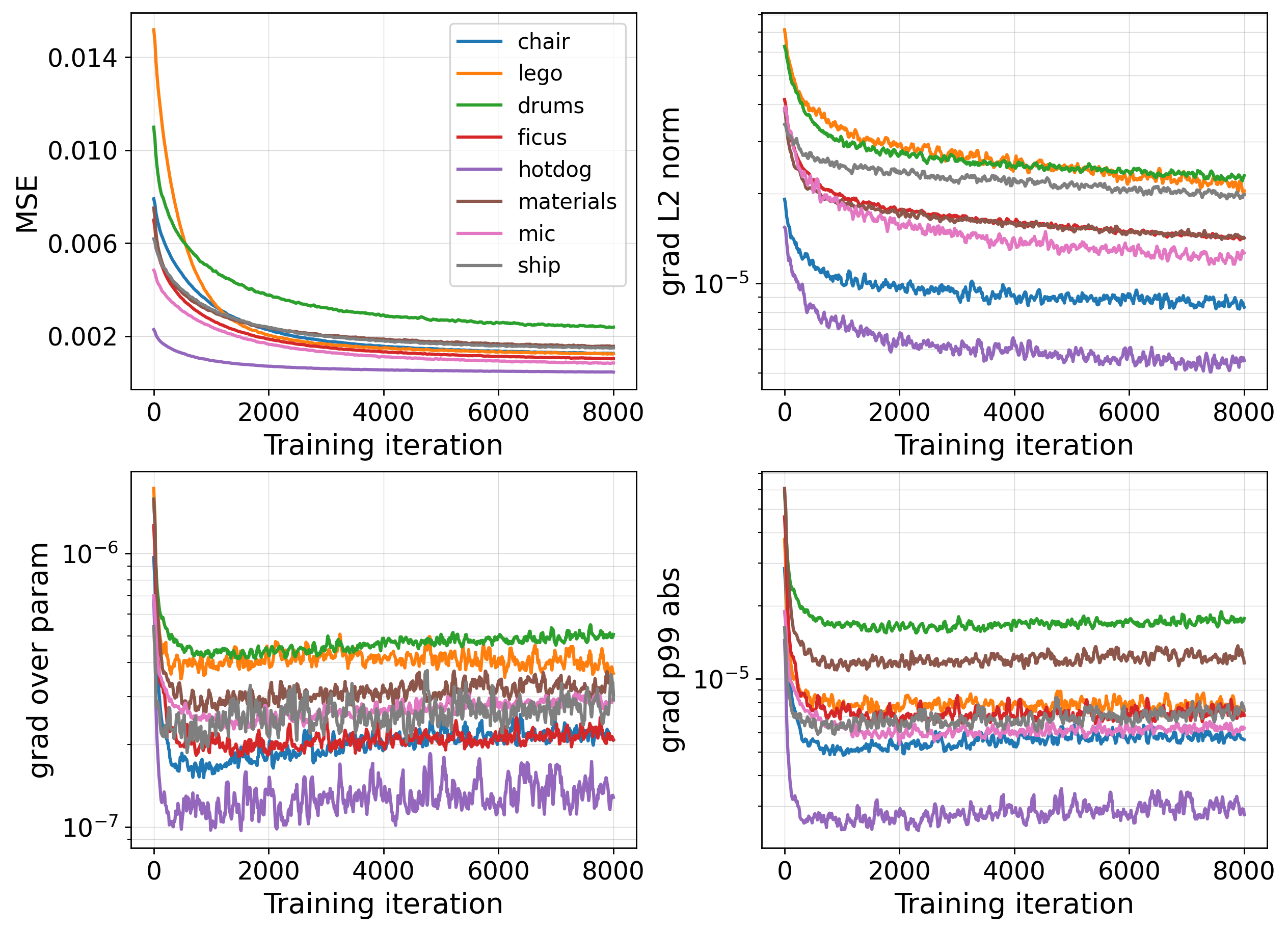}
   \caption{
    Training diagnostics of \ours{} with STE as the gradient surrogate through the codec round trip. We plot the reconstruction MSE, gradient $\ell_2$ norm of the feature plane parameters, the scale-normalized gradient magnitude, and the 99th-percentile absolute gradient magnitude (computed from 200k sampled parameters). All metrics remain bounded while the loss decreases, indicating stable optimization without exploding or vanishing gradients.
    }
   \label{fig:gradient_diagnostics}
\end{figure}

\begin{figure*}[t]
  \centering

   \includegraphics[width=1.0\linewidth]{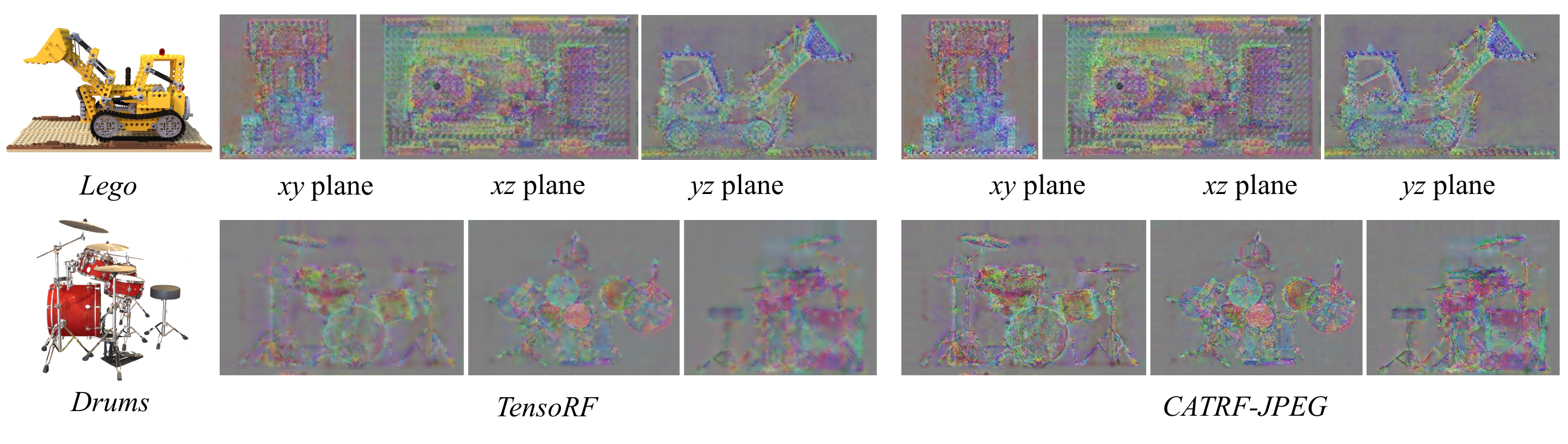}

   \caption{
    Comparison of visualized appearance planes packed with \textit{FlattenRGB}. For a given scene, we show the $xy$, $xz$, and $yz$ planes of vanilla TensoRF (left) and our \ours{}-JPEG model (right). The CATRF canvases exhibit richer high-frequency structure than the vanilla TensoRF planes, reflecting that our SCL training explicitly adapts feature planes to the codec round trip. Because the codec’s transform, quantization, and in-loop filtering tend to smooth the canvases, CATRF learns to encode more detail pre-encoding. Consequently, after decoding, the decoded patterns still preserve the underlying feature semantics needed for high-quality rendering.
    }
   \label{fig:canvas_comparison}
\end{figure*}

\paragraph{Observed payload for CNC.}
For CNC~\cite{chen2024far}, we directly reuse the bitrate estimation utilities provided in their official implementation, which have already  included quantization and entropy coding. The difference between the sizes reported in our evaluation and those in the original paper stems from whether one includes the compressed MLP sizes and entropy model parameters in the total bitrate budget.

\section{More Implementation Details}
For the radiance-field backbone, we train for 30k iterations and adopt progressive scaling widely used in the literature. The plane resolutions (and line vectors) are bilinearly upsampled at steps [2000, 3000, 4000, 5500, 7000], and the alpha-mask is recomputed on a schedule [2000, 4000, 6000, 11000, 16000, 21000, 26000]. The MLP render has two hidden layers (128 units, ReLU) and a 3-channel head with sigmoid. 
Optimization uses Adam with learning rates $lr=0.02$ for the radiance-field model and $lr=0.001$ for the renderer MLP. 
When employing AbsMax (absolute max) quantization, we employ hard-coded upper and lower bounds. For the static branch (TensoRF backbone), density and appearance planes are globally quantized within preset ranges, $[-25, 25]$ and $[-5, 5]$, respectively. For the dynamic branch (DVGO backbone), density and appearance planes are both globally quantized within the preset range, $[-20, 20]$.
For our STE-based caching, the change-based refresh uses a threshold $\epsilon=0.05$ ($5\%$ normalized $\ell_2$ drift) to trigger early codec refreshes. 

\section{Gradient Diagnostics}
The core of \ours{} is treating the entire codec encoding-decoding round trip as a black box and using a Straight-Through Estimator (STE) as its gradient surrogate. 
While this enables backpropagation, the resulting gradients are biased: standard codecs introduce non-linear distortions that an identity STE does not explicitly model, potentially causing gradient mismatch and training instability. 
Given the complexity of standard codecs (e.g., quantization, block transforms, entropy coding, and in-loop filtering), providing a formal convergence proof is intractable. 
Instead, we provide empirical evidence that STE yields stable and useful surrogate gradients in our training setup. Specifically, we log four statistics throughout training:
(1) \textit{MSE}, the reconstruction loss between rendered RGB and ground-truth images; 
(2) \textit{grad L2 norm}, the $\ell_2$ norm of gradients on the feature-plane parameters; 
(3) \textit{grad over param}, a scale-normalized gradient magnitude $\|\nabla_\theta\mathcal{L}\|_2 / (\|\theta\|_2+\epsilon)$; and
(4) \textit{grad p99 abs}, the 99th percentile of $|\nabla_\theta\mathcal{L}|$ computed from a random sample of 200{,}000 gradient entries.
As shown in \cref{fig:gradient_diagnostics}, these metrics remain bounded (with no NaN/Inf entries) while the MSE decreases smoothly, indicating that optimization proceeds normally under STE despite the biased gradients. 
This diagnostic supports our design choice that an STE-through-codec surrogate is sufficient for stable SCL training.


\section{Gradient-Surrogate Ablation}
Surrogate-gradient design for non-differentiable operators is a long-standing problem in quantized and discrete optimization. 
Beyond the vanilla identity Straight-Through Estimator (STE), many alternatives have been proposed to reduce gradient bias or better approximate the local effect of quantization. 
Since \ours{} treats the full codec round trip as a black box, it is natural to ask whether a more sophisticated surrogate than vanilla STE would further improve optimization. 
We therefore compare three strategies: vanilla STE~\cite{bengio2013estimating}, modified STE (mSTE)~\cite{mack2025efficient}, and a hybrid STE+SPSA estimator inspired by black-box gradient approximation~\cite{spall1998overview,ramirez2021differentiable}. 
\cref{tab:grad_surrogate_ablation} reports rate--distortion performance (PSNR/SSIM/LPIPS), compressed size, and training time.
Vanilla STE uses the identity surrogate in the backward pass, i.e., the gradient is copied directly to the pre-codec feature plane. 
Although this surrogate is biased, recent work has noted that STE can nevertheless provide high-quality biased gradients in practice~\cite{yang2025improving}. 
Our results are consistent with this observation. Vanilla STE provides the most robust optimization overall, particularly in the low-bitrate regime where codec-induced distortions are strongest. 
We therefore use vanilla STE as the default gradient surrogate in all main experiments.
The modified STE (mSTE) rescales the detached codec error using the standard deviation of the quantization error. 
Let $\epsilon = \hat{\mathbf{P}} - \mathbf{P}$ denote the codec-induced perturbation between the decoded feature plane $\hat{\mathbf{P}}$ and the pre-codec feature plane $\mathbf{P}$. 
mSTE takes the form
\[
\widetilde{\mathbf{P}}
=
\mathbf{P}
+
\operatorname{sg}(\epsilon)
\cdot
\frac{\sigma(\epsilon)}{\operatorname{sg}(\sigma(\epsilon))},
\]
where $\operatorname{sg}(\cdot)$ is the stop-gradient operator and $\sigma(\epsilon)$ is the standard deviation of the perturbation. 
Intuitively, this preserves the STE forward pass while modulating the backward signal using the observed perturbation scale. 
In our experiments, mSTE performs competitively at high bitrate, but degrades noticeably at low bitrate. 
A plausible reason is that, under stronger codec distortion, the perturbation becomes less like a small local quantization residual and more strongly reflects the non-linear artifacts from the black-box codec. Thus, a standard-deviation reweighting can become a poor proxy for the true local sensitivity instead, making optimization less stable and less effective than plain identity STE.
We also evaluate SPSA, a gradient estimator based on symmetric random perturbations~\cite{spall1998overview}. 
For an objective $C_q(.)$, SPSA estimates its gradient by running two function evaluations:
\[
\hat{g}_i(\mathbf{P})
=
\frac{
\mathcal{C}_q(\mathbf{P} + \epsilon \Delta) - \mathcal{C}_q(\mathbf{P} - \epsilon \Delta)
}{
2 \epsilon \Delta_i
},
\]
where $\Delta$ is a random perturbation vector sampled from a symmetric Bernoulli distribution, i.e., $\Delta_i=\pm 1$ with probability of 0.5; $\epsilon$ is a small perturbation scale. 
In our implementation, we use a hybrid STE+SPSA scheme, where STE is used on cached steps, while SPSA is applied only when the codec cache is refreshed. 
This design keeps the method computationally feasible, since applying SPSA at every step would require multiple additional codec round trips per iteration and would be prohibitively expensive in our setting. 
Empirically, STE+SPSA yields the worst reconstruction quality and the highest training time. 
We attribute this to the high variance of zeroth-order estimates and the poor signal-to-noise ratio introduced by severe codec non-linearity. Although SPSA is less biased in principle, its gradient estimates can be substantially noisier for a complex, black-box codec roundtrip, making them less aligned with the descent direction needed for radiance-field optimization.
Overall, this ablation shows that reducing bias in the surrogate gradient does not necessarily improve optimization for SCL TriPlane training. 
Instead, vanilla STE offers the best trade-off between stability, RD performance, and efficiency. 

\begin{table}[t!]
    \setlength{\tabcolsep}{4pt}  
    \centering
    \resizebox{\linewidth}{!}{
    \begin{tabular}{cccccc}
    \toprule
    & PSNR & SSIM & LPIPS & Size & Training \\
    \midrule
    STE (high-rate)~\cite{bengio2013estimating} & 33.56 & 0.969 & 0.182 & 0.27 MB & 1h 14m\\
    STE (low-rate)~\cite{bengio2013estimating} & 31.91 & 0.957 & 0.267 & 0.13 MB & 1h 9m\\

    mSTE (high-rate)~\cite{mack2025efficient} & 33.62 & 0.968 & 0.195 & 0.32 MB & 1h 25m\\
    mSTE (low-rate)~\cite{mack2025efficient} & 30.96 & 0.948  & 0.368 & 0.17 MB & 1h 28m\\

    STE+SPSA (high-rate)~\cite{spall1998overview} & 29.63 & 0.930 & 0.459 & 0.14 MB & 3h 18m\\
    STE+SPSA (low-rate)~\cite{spall1998overview} & 27.94 & 0.909 & 0.650 & 0.07 MB& 3h 18m\\
    \bottomrule
    \end{tabular}
    }
    \caption{
    Ablation of gradient-surrogate methods for SCL training. STE demonstrates superior robustness at low rates, whereas mSTE degrades under heavy quantization. STE+SPSA yields suboptimal R-D performance and incurs substantial training latency due to repeated codec round-trips.
    }
    \label{tab:grad_surrogate_ablation}
\end{table}

\section{More Quantitative and Qualitative Results}
\label{sec:more_results}

\begin{figure*}[t]
  \centering

   \includegraphics[width=1.0\linewidth]{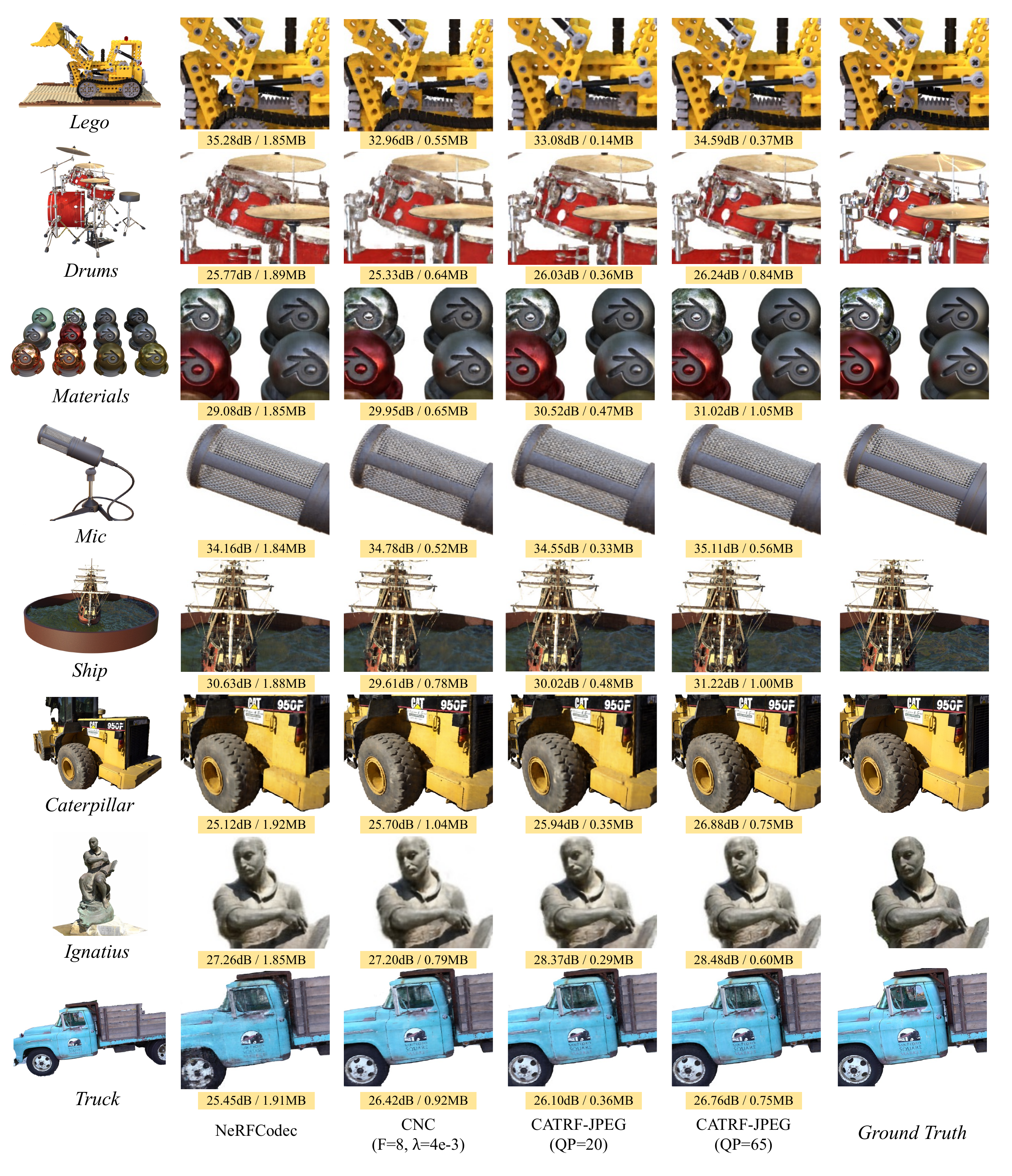}

   \caption{More qualitative comparisons of NeRF Synthetic and Tanks and Temples benchmarks.
   }
   \label{fig:more_qual_supp}
\end{figure*}

\begin{figure*}[t]
  \centering

   \includegraphics[width=0.85\linewidth]{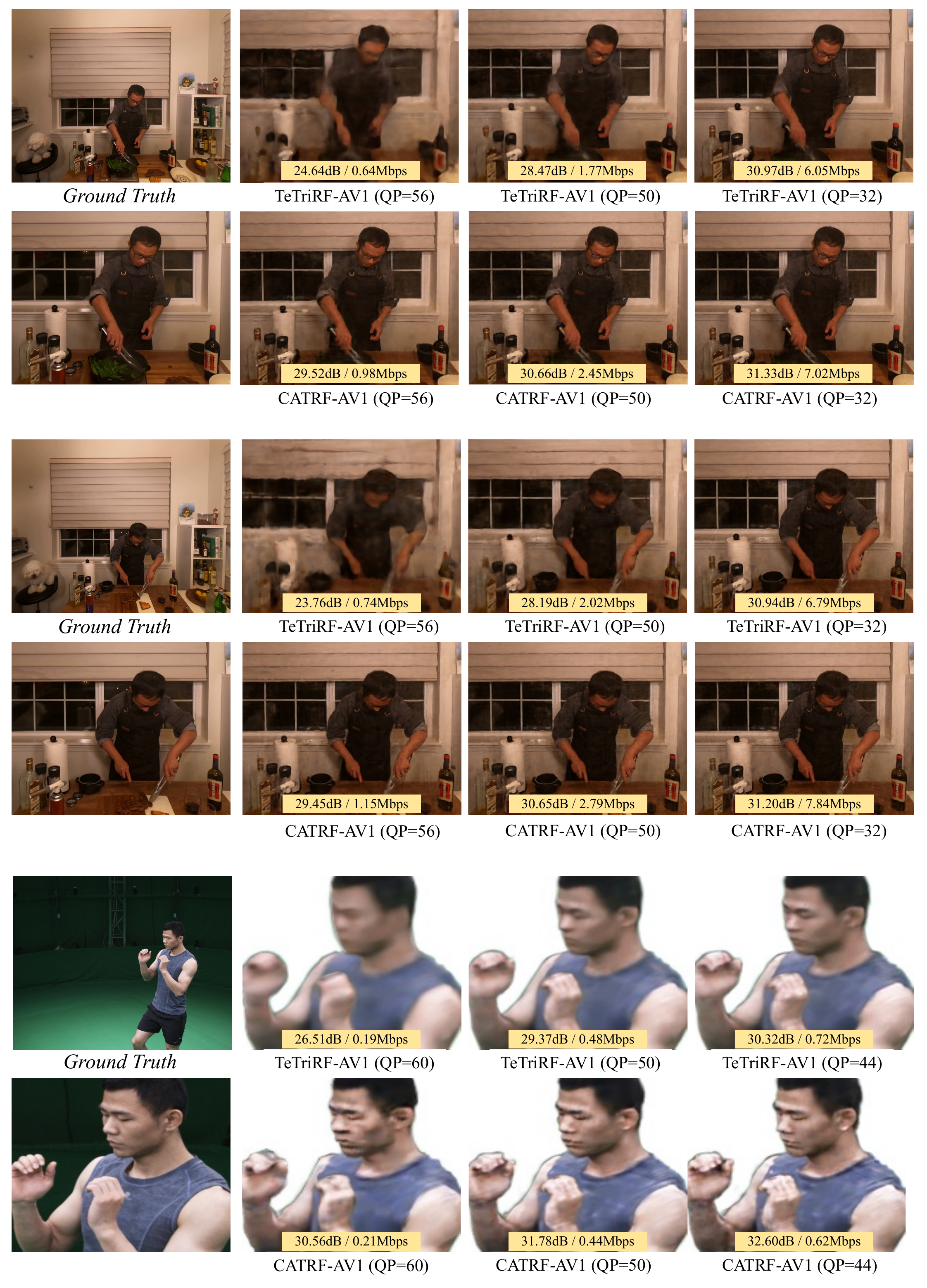}

   \caption{More qualitative comparisons of Neural 3D Video and NHR benchmarks.
   }
   \label{fig:more_video_qual_supp}
\end{figure*}

In this section, we present detailed quantitative and qualitative results that complement the main paper. First, we report per-scene quantitative results for the NeRF Synthetic and Tanks and Temples benchmarks, allowing a more fine-grained comparison across individual scenes (see Tab.~\ref{tab:per_scene_nerf} and Tab.~\ref{tab:per_scene_tank}). We then provide extra qualitative visualizations for both static and dynamic benchmarks to further illustrate how \ours{} behaves across a wide range of operating points and scene types (see Fig.~\ref{fig:more_qual_supp} and Fig.~\ref{fig:more_video_qual_supp}).

\begin{table*}[t!]
    \centering
    \resizebox{\linewidth}{!}{
    \begin{tabular}{cccccccccc}
    \toprule
    Method & chair & drums & ficus & hotdog & lego & materials & mic & ship & Avg. \\
    \hline
    \multicolumn{10}{c}{{PSNR}} \\
    \hline
    TensoRF~\cite{chen2022tensorf} & 35.89 & 26.39 & 34.34& 36.76 & 34.98& 30.96& 34.68& 30.43& 33.05\\
    NeRFCodec~\cite{li2024nerfcodec} & 34.28 & 25.77& 33.25& 35.07& 35.28& 29.08& 34.16& 30.63& 32.19\\
    CNC~\cite{chen2024far} ($F=8$, $\lambda=4e-3$) & 32.99 & 25.33 & 32.31 & 35.66 & 32.96 & 29.95 & 34.78 & 29.61 & 31.70\\
    CNC~\cite{chen2024far} ($F=8$, $\lambda=2e-3$) & 34.45 & 25.43 & 33.27 & 35.92 & 34.50 & 30.06 & 36.62 & 29.65 & 32.49\\
    CNC~\cite{chen2024far} ($F=8$, $\lambda=7e-4$) & 35.56 & 26.18 & 32.88 & 37.39 & 34.64 & 30.27 & 36.59 & 31.83 & 33.16\\
    
    HAC++~\cite{chen2025hac++} ($\lambda=3e-3$) & 34.42 & 26.28 & 34.46 & 36.76 & 34.36 & 29.99 & 34.90 & 30.49 & 32.71 \\
    HAC++~\cite{chen2025hac++} ($\lambda=2e-3$) & 34.89 & 26.34 & 34.55 & 37.00 & 34.71 & 30.21 & 35.54 & 30.70 & 32.99 \\
    HAC++~\cite{chen2025hac++} ($\lambda=1e-3$) & 35.55 & 26.49 & 34.68 & 37.52 & 35.10 & 30.33 & 35.80 & 30.70 & 33.27 \\
    
    \ours{}-JPEG (QP=20) & 34.74& 26.03 & 33.75 & 36.15& 33.08& 30.52& 34.55& 30.02&32.17\\
    \ours{}-JPEG (QP=35) & 35.06& 26.10 & 34.18 & 36.38& 33.61& 30.78& 34.77& 30.77& 32.49\\
    \ours{}-JPEG (QP=65) & 35.79& 26.24 & 34.43 & 36.71& 34.59 & 31.02 & 35.11& 31.22& 33.13\\
    \hline
    \multicolumn{10}{c}{{SSIM}} \\
    \hline
    TensoRF~\cite{chen2022tensorf} & 0.983 & 0.931 & 0.983& 0.981 & 0.978& 0.957& 0.988& 0.888& 0.961\\
    NeRFCodec~\cite{li2024nerfcodec} & 0.976 & 0.923& 0.978& 0.968& 0.974& 0.950& 0.988& 0.877& 0.954\\
    CNC~\cite{chen2024far} ($F=8$, $\lambda=4e-3$) & 0.980 & 0.941 & 0.983 & 0.978 & 0.978 & 0.958 & 0.991 & 0.901 & 0.964\\
    CNC~\cite{chen2024far} ($F=8$, $\lambda=2e-3$) &  0.982 & 0.942 & 0.984 & 0.980 & 0.980 & 0.959 & 0.992 & 0.909 & 0.966\\
    CNC~\cite{chen2024far} ($F=8$, $\lambda=7e-4$) & 0.984 & 0.942 & 0.984 & 0.982 & 0.982 & 0.960 & 0.993 & 0.915 & 0.968\\

    HAC++~\cite{chen2025hac++} ($\lambda=3e-3$) & 0.980 & 0.951 & 0.985 & 0.981 & 0.977 & 0.961 & 0.989 & 0.902 & 0.966\\
    HAC++~\cite{chen2025hac++} ($\lambda=2e-3$) & 0.982 & 0.951 & 0.986 & 0.982 & 0.978 & 0.963 & 0.991 & 0.903 & 0.967 \\
    HAC++~\cite{chen2025hac++} ($\lambda=1e-3$) & 0.985 & 0.952 & 0.986 & 0.983 & 0.980 & 0.964 & 0.991 & 0.903 & 0.968\\
    
    \ours{}-JPEG (QP=20) & 0.974& 0.925 & 0.977 & 0.973& 0.958& 0.954& 0.983& 0.874&0.952\\
    \ours{}-JPEG (QP=35) & 0.978& 0.928 & 0.979 & 0.975& 0.963& 0.957& 0.985& 0.881&0.956\\
    \ours{}-JPEG (QP=65) & 0.981& 0.930 & 0.982 & 0.978& 0.970& 0.970& 0.987& 0.886&0.961\\
    \hline
    \multicolumn{10}{c}{{Size (MB)}} \\
    \hline
    TensoRF~\cite{chen2022tensorf} & 69.19 & 69.26 & 72.02 & 85.24 & 17.99 & 86.14 & 67.96 & 72.24 & 67.51\\

    NeRFCodec$^\text{\ding{61}}$~\cite{li2024nerfcodec} & 0.525 & 0.602 & 0.563 & 0.521& 0.555& 0.556& 0.553& 0.593& 0.559\\
    NeRFCodec~\cite{li2024nerfcodec} & 1.819 & 1.896 & 1.863 & 1.721& 1.849& 1.850& 1.847& 1.887& 1.842\\ 

    CNC~\cite{chen2024far} ($F=8$, $\lambda=4e-3$)$^\text{\ding{61}}$ & 0.406 &0.488 &0.365 &0.332 &0.377 &0.485 &0.332 &0.560 &0.418  \\
    CNC~\cite{chen2024far} ($F=8$, $\lambda=2e-3$)$^\text{\ding{61}}$ & 0.511 &0.649 &0.444 &0.367 &0.454 &0.610 &0.366 &0.717 &0.515  \\
    CNC~\cite{chen2024far} ($F=8$, $\lambda=7e-4$)$^\text{\ding{61}}$ & 0.689 &1.003 &0.588 &0.470 &0.602 &0.851 &0.471 &1.106 &0.722  \\

    CNC~\cite{chen2024far} ($F=8$, $\lambda=4e-3$)
      & 0.652 & 0.641 & 0.595 & 0.518 & 0.545 & 0.647 & 0.524 & 0.780 & 0.613 \\
    CNC~\cite{chen2024far} ($F=8$, $\lambda=2e-3$)
      & 0.668 & 0.802 & 0.622 & 0.553 & 0.622 & 0.772 & 0.744 & 0.938 & 0.715 \\
    CNC~\cite{chen2024far} ($F=8$, $\lambda=7e-4$)
      & 0.847 & 1.130 & 0.703 & 0.656 & 0.770 & 1.013 & 1.486 & 1.327 & 0.991 \\

    HAC++~\cite{chen2025hac++} ($\lambda=3e-3$) & 0.707 & 1.220 & 0.766 & 0.620 & 0.892 & 1.027 & 0.574 & 1.256 & 0.883 \\
    HAC++~\cite{chen2025hac++} ($\lambda=2e-3$) & 0.830 & 1.546 & 0.917 & 0.721 & 1.223 & 1.166 & 0.654 & 1.541 & 1.075\\
    HAC++~\cite{chen2025hac++} ($\lambda=1e-3$) & 1.098 & 1.787 & 1.219 & 0.802 & 1.434 & 1.498 & 0.868 & 2.150 & 1.357\\
   
    \ours{}-JPEG (QP=20) & 0.332 & 0.364& 0.368& 0.362& 0.138& 0.473& 0.335& 0.481& 0.357\\
    \ours{}-JPEG (QP=35) & 0.453 & 0.509& 0.501& 0.477& 0.186& 0.636& 0.408& 0.664&0.479\\
    \ours{}-JPEG (QP=65) & 0.731 & 0.840& 0.797& 0.767& 0.370& 1.049& 0.559& 1.001&0.764\\
    \bottomrule
    \end{tabular}
    }
    \caption{
    Per-scene quantitative results on the NeRF Synthetic dataset~\cite{mildenhall2021nerf}. When computing the model size, we include all scene-specific codec and entropy-model parameters required to decode the radiance fields at the client. We also report the size of the encoded radiance fields alone, indicated by $\text{\ding{61}}$. For per-scene results of other baselines, please refer to ECRF~\cite{lee2024ecrf}.
    }
    \label{tab:per_scene_nerf}
\end{table*}

\begin{table*}[t!]
    \centering
    \normalsize            
    \setlength{\tabcolsep}{4pt}  
    \begin{tabular}{ccccccc}
    \toprule
    Method & Barn & Caterpillar & Family & Ignatius & Truck  & Avg. \\
    \hline
    \multicolumn{7}{c}{{PSNR}} \\
    \hline
    TensoRF~\cite{chen2022tensorf} & 29.48 & 26.84 & 33.69 & 28.64 & 26.88 & 29.11\\
    NeRFCodec~\cite{li2024nerfcodec} & 28.35 & 25.12 & 33.39& 27.26 & 25.45&  27.91\\
    CNC ($F=8$, $\lambda=8e-3$)~\cite{chen2024far} & 28.15 & 25.85 & 32.48 & 27.52 & 26.38 &  28.08\\
    CNC ($F=8$, $\lambda=4e-3$)~\cite{chen2024far} & 28.56 & 25.70 & 32.68 & 27.20 & 26.42 & 28.11\\
    CNC ($F=8$, $\lambda=2e-3$)~\cite{chen2024far} & 28.75 & 26.22 & 32.72 & 27.53 & 26.23 & 28.29\\
    CNC ($F=8$, $\lambda=7e-4$)~\cite{chen2024far} & 28.82 & 26.44 & 32.86 & 28.02 & 27.12 & 28.65\\

    \ours{}-JPEG (QP=20) & 28.69& 25.94 & 31.86 & 28.37& 26.10& 28.19\\
    \ours{}-JPEG (QP=35) & 29.02& 26.30 & 32.21 & 28.45& 26.44& 28.48\\
    \ours{}-JPEG (QP=50) & 29.45& 26.86 & 33.42 & 28.52& 26.60& 28.97\\
    \ours{}-JPEG (QP=65) & 29.67& 26.88 & 33.72 & 28.48& 26.76 & 29.10\\
    \hline
    \multicolumn{7}{c}{{SSIM}} \\
    \hline
    TensoRF~\cite{chen2022tensorf} & 0.901 & 0.913 & 0.965 & 0.949 & 0.903 & 0.926\\
    NeRFCodec~\cite{li2024nerfcodec} & 0.849 & 0.891 & 0.957& 0.940 & 0.863&  0.901\\
    CNC ($F=8$, $\lambda=8e-3$)~\cite{chen2024far} & 0.866 &0.911 &0.955 &0.941 &0.910 &0.917\\
    CNC ($F=8$, $\lambda=4e-3$)~\cite{chen2024far} & 0.872 &0.914 &0.959 &0.944 &0.914 &0.921\\
    CNC ($F=8$, $\lambda=2e-3$)~\cite{chen2024far} & 0.879 &0.917 &0.961 &0.946 &0.917 &0.924\\
    CNC ($F=8$, $\lambda=7e-4$)~\cite{chen2024far} & 0.884 &0.920 &0.965 &0.947 &0.921 &0.927\\

    \ours{}-JPEG (QP=20) & 0.893 & 0.895& 0.968 & 0.946 & 0.892&  0.919\\
    \ours{}-JPEG (QP=35) & 0.904 & 0.905& 0.971 & 0.949 & 0.900& 0.926\\
    \ours{}-JPEG (QP=50) & 0.909 & 0.908& 0.973 & 0.951 & 0.904& 0.929\\
    \ours{}-JPEG (QP=65) & 0.923 & 0.918& 0.976 & 0.956 & 0.916& 0.938\\
    
    \hline
    \multicolumn{7}{c}{{Size (MB)}} \\
    \hline
    TensoRF~\cite{chen2022tensorf} & 82.06 & 73.22 & 68.42 & 68.18 & 78.52 & 74.08\\
    
    NeRFCodec$^\text{\ding{61}}$~\cite{li2024nerfcodec} & 0.574 & 0.625 & 0.561 & 0.552& 0.613& 0.584\\
    NeRFCodec~\cite{li2024nerfcodec} & 1.867 & 1.925 & 1.861 & 1.852& 1.913& 1.884\\
    
    CNC ($F=8$, $\lambda=8e-3$)$^\text{\ding{61}}$~\cite{chen2024far} & 0.546 & 0.579 & 0.384 & 0.432 & 0.511 & 0.490\\
    CNC ($F=8$, $\lambda=4e-3$)$^\text{\ding{61}}$~\cite{chen2024far} & 0.726 & 0.824 & 0.455 & 0.559 & 0.708 & 0.654 \\
    CNC ($F=8$, $\lambda=2e-3$)$^\text{\ding{61}}$~\cite{chen2024far} & 0.976 & 1.067 & 0.543 & 0.721 & 0.992 & 0.860\\
    CNC ($F=8$, $\lambda=7e-4$)$^\text{\ding{61}}$~\cite{chen2024far} & 1.465 & 1.652 & 0.710 & 1.146 & 1.539 & 1.302\\

    CNC ($F=8$, $\lambda=8e-3$)~\cite{chen2024far} 
      & 0.807 & 0.799 & 0.598 & 0.671 & 0.726 & 0.720\\
    CNC ($F=8$, $\lambda=4e-3$)~\cite{chen2024far} 
      & 0.987 & 1.044 & 0.669 & 0.798 & 0.923 & 0.884\\
    CNC ($F=8$, $\lambda=2e-3$)~\cite{chen2024far} 
      & 1.237 & 1.287 & 0.757 & 0.960 & 1.207 & 1.090\\
    CNC ($F=8$, $\lambda=7e-4$)~\cite{chen2024far}
      & 1.726 & 1.872 & 0.924 & 1.385 & 1.754 & 1.532\\
      
    \ours{}-JPEG (QP=20) & 0.368 & 0.356& 0.501& 0.296& 0.365& 0.377\\
    \ours{}-JPEG (QP=35) & 0.504 & 0.483& 0.671& 0.393& 0.486& 0.507\\
    \ours{}-JPEG (QP=50) & 0.675 & 0.626& 0.851& 0.501& 0.634& 0.657\\
    \ours{}-JPEG (QP=65) & 0.816 & 0.746& 1.008& 0.599& 0.755& 0.785\\
    \bottomrule
    \end{tabular}
    
    \caption{Per-scene results on the Tanks and Temples dataset~\cite{liu2020neural}. When computing the model size, we include all scene-specific codec and entropy-model parameters required to decode the radiance fields at the client. We also report the size of the encoded radiance fields alone, indicated by $\text{\ding{61}}$. For per-scene results of other baselines, please refer to ECRF~\cite{lee2024ecrf}}
    \label{tab:per_scene_tank}
\end{table*}

\end{document}